\documentclass[11pt]{article}
\usepackage{amsmath,amssymb,color,graphics,epsfig,cite}
\usepackage{amsfonts}
\newcommand{\hoch}[1]{$\, ^{#1}$}

\newcommand{\be}{\begin{equation}}
\newcommand{\ee}{\end{equation}}
\newcommand{\bea}{\setlength\arraycolsep{2pt} \begin{eqnarray}}
\newcommand{\eea}{\end{eqnarray}}

\def\0{{\sst{(0)}}}
\def\1{{\sst{(1)}}}
\def\2{{\sst{(2)}}}
\def\3{{\sst{(3)}}}
\def\4{{\sst{(4)}}}
\def\5{{\sst{(5)}}}
\def\6{{\sst{(6)}}}
\def\7{{\sst{(7)}}}
\def\8{{\sst{(8)}}}
\def\sst#1{{\scriptscriptstyle #1}}

\begin{document}

\begin{center}
{\Large {\bf Cubic Curvature Corrections to Taub-NUT-AdS Black Holes and Their Thermodynamics }}

\vspace{20pt}

{\large Yu-Qi Chen\hoch{1}, Hai-Shan Liu\hoch{1} and H. L\"u\hoch{1,2}}

\vspace{10pt}

{\it \hoch{1}Center for Joint Quantum Studies, Department of Physics,\\
School of Science, Tianjin University, Tianjin 300350, China }

\vspace{5pt}

{\it \hoch{2}The International Joint Institute of Tianjin University, Fuzhou,\\ Tianjin University, Tianjin 300350, China}

\vspace{40pt}

\underline{ABSTRACT}

\end{center}

We investigate cubic curvature corrections to four-dimensional Taub-NUT-AdS black holes and their thermodynamics. We construct the complete analytic solution to the first order in the higher-derivative couplings, and derive the corrected Hawking temperature and Wald entropy. The NUT potential remains unchanged, while reconstructing the NUT charge and Gibbs free energy through the Maxwell relation leaves just one undetermined function of the NUT parameter. We then independently evaluate the Gibbs free energy using two improved Reall-Santos methods, based on either the field redefinition or the background subtraction. After matching the thermodynamic variables, both methods yield identical results and fix the undetermined function, providing a consistent determination of the mass and NUT charge. Our work therefore establishes that the thermodynamics of Taub-NUT-AdS black hole withstands the higher-derivative corrections.

\vfill{\footnotesize yuqi\_chen@tju.edu.cn\ \ \ hsliu.zju@gmail.com\ \ \ mrhonglu@gmail.com }
%\vfill {\footnotesize mrhonglu@gmail.com}

%{\footnotesize \hoch{*}Corresponding author}

\thispagestyle{empty}
\pagebreak
%\voffset=-40pt
%\setcounter{page}{1}

%\tableofcontents
%\addtocontents{toc}{\protect\setcounter{tocdepth}{2}}

%\newpage

\section{Introduction}

One of the most important developments in theoretical physics over the past few decades is the AdS/CFT correspondence \cite{Maldacena:1997re,Witten:1998qj, Gubser:1998bc}. In this framework, quantum gravity in an anti-de Sitter (AdS) spacetime is conjectured to be dual to a conformal field theory living on the AdS boundary. This correspondence has since inspired the construction of new exact solutions of asymptotically AdS spacetimes. Within the framework of pure gravity, the inclusion of a cosmological constant in the Schwarzschild black hole in general dimensions is straightforward; however, adding a cosmological constant to the Kerr black hole \cite{Kerr:1963ud} is much more subtle, and was achieved by Carter, motivated at the time not physically but by a geometric curiosity \cite{Carter:1968ks}. The AdS/CFT correspondence provides such a physical motivation, and consequently, the cosmological constant was added to the Myers-Perry black holes \cite{Myers:1986un}, first in five dimensions \cite{Hawking:1998kw} and then in general dimensions \cite{Gibbons:2004uw, Gibbons:2004js}. Furthermore, the Plebanski metric \cite{Plebanski:1975xfb} can also be generalized to higher dimensions \cite{Chen:2006xh}. There are also less successful examples of the generalizations. A notable one is that the Plebanski-Demianski solutions \cite{Plebanski:1976gy} are generalized so far only to $D=5$ without a cosmological constant \cite{Lu:2008js}. There has been no success yet to introduce a cosmological constant to the recently constructed Ricci-flat $B$-deformed Schwarzschild black hole \cite{Astorino:2026okd} or the two $B$-deformed Kerr black holes \cite{Ma:2026otg, Ma:2026gdn}.

The Taub-NUT black hole \cite{Taub:1950ez,Newman:1963yy} is another remarkable exact solution in general relativity, constructed in 50s among many important emerging solutions at the time. Compared to the Schwarzschild metric of mass $m$, the Taub-NUT geometry contains an additional parameter, typically denoted as $n$ and referred to as the NUT parameter. The Ricci-flat metric is of cohomogeneity one, with level surfaces of a time-bundle over round two-spheres, and $n$ parameterizes the fibre. The presence of such a NUT parameter changes the asymptotic structure of the spacetime, such that it is no longer asymptotic to Minkowski, but asymptotically locally Minkowskian. Moreover, the spacetime contains a global singular structure located along the north and south axes, known as the Misner string \cite{Misner:1963fr}. Nevertheless, Taub-NUT spacetime can still be geodesically complete when the Misner string is appropriately treated \cite{Clement:2015cxa,Clement:2015aka}.

The recent interest in the Taub-NUT spacetime partly arises from the astrophysical observation that our world might have NUTty influence \cite{Chakraborty:2017nfu, Ghasemi-Nodehi:2021ipd,Chakraborty:2022ltc}. The main reason is geometric curiosity about its global structure, particularly on whether we can establish a sensible black hole thermodynamic system in the presence of a Misner string. Since the initial work of \cite{Hennigar:2019ive}, there has been extensive research on the subject with many different proposals \cite{Bordo:2019slw,BallonBordo:2019vrn,Durka:2019ajz, Wu:2019pzr,Chen:2019uhp,BallonBordo:2020mcs,Awad:2020dhy,Abbasvandi:2021nyv,Frodden:2021ces, Wu:2022rmx,Godazgar:2022jxm,Awad:2022jgn,Wu:2022mlz,Wu:2022xpp,Wu:2023woq,Liu:2022wku,
Liu:2023uqf}. There are many reasons for the subtleties that lead to quite a few proposals. Although most of the proposals have the same temperature $T$ and entropy $S$, they differ from the definitions of mass $M$ and NUT charge $Q_n$. The disagreement on the former originates from the fact that the solution is only asymptotically locally Minkowskian spacetime and hence the standard definition of mass no longer exists. The NUT charge is also a controversial concept and has no universally-accepted definition. The Iyer-Wald formalism \cite{Wald:1993nt,Iyer:1994ys} guarantees the existence of a first-order differential identity among the integration constants involving a term of $TdS$. However, it requires a careful treatment of excluding the Misner strings \cite{Liu:2022wku}. Moreover, the Iyer-Wald formalism does not by itself tells one what is mass and the NUT charge. This leads to the situation that different proposals of mass and NUT charge can all satisfy the first law. Remarkably, many of the proposals also passed the test of the Euclidean action approach \cite{Gibbons:1976ue}.

In this paper, we adopt the black hole thermodynamics derived from the formalism proposed in \cite{Liu:2022wku}, where the mass and the NUT charge can be independently obtained from the appropriate Komar integration of the closed 2-forms associated with the Killing vectors of time and azimuthal angle coordinates respectively, while the conjugate NUT potential, $\Phi_n=n/2$, is solely determined by the fibre bundle structure. The presence of the Misner string makes the evaluation of the Komar integration rather subtle, but the end result is consistent with both the Iyer-Wald first law and Euclidean action principle. The treatment was also generalized to the Taub-NUT-Kerr black hole \cite{Liu:2022wku}.

The inclusion of a cosmological constant and the construction of the Taub-NUT-AdS is straightforward \cite{Hawking:1998ct}, and the thermodynamic frameworks have also been applied in holographic studies \cite{Jiang:2019yzs,Chen:2023eio,Perry:2022udk, Siahaan:2022jrl}. In this paper, we study the higher-derivative corrections to the Taub-NUT-AdS black hole. It is widely accepted that Einstein gravity itself is not a
complete quantum theory because it is perturbatively non-renormalizable
\cite{tHooft:1974toh}. From the perspective of effective field theory,
one can systematically include all possible diffeomorphism-invariant
higher-curvature corrections to the Einstein-Hilbert action, leading to a
low-energy effective gravitational theory,
\begin{equation}
	I=\frac{1}{16\pi}\int d^4x\sqrt{-g}
	\left(R-2\Lambda+\mathcal{L}_{\rm HD}\right)
	+I_{\rm surf}+I_{\rm ct},
\end{equation}
where $\mathcal{L}_{\rm HD}$ contains higher-derivative curvature corrections,
\begin{equation}
	\mathcal{L}_{\rm HD}
	=
	L^2\mathcal{L}_{4\partial}
	+L^4\mathcal{L}_{6\partial}
	+\cdots .
\end{equation}
Here $I_{\rm surf}$ denotes the boundary contribution for the consistency of the Hamilton's variation principle, including the Gibbons-Hawking-York (GHY) associated with the Einstein-Hilbert term, as well as the terms required by the higher-curvature corrections. The term $I_{\rm ct}$ represents the holographic counterterms introduced to remove the ultraviolet divergences of asymptotically AdS spacetimes.

However, the boundary terms play no role in the construction of the local solution.  Although the Taub-NUT geometry is not static, it is of cohomogeneity one, with a time bundle over the round two-sphere, so that the metric is spherically-symmetric in that it is invariant under the transitive action on the two-sphere. It is thus consistent to propose an spherical-symmetric ansatz that preserves the time-bundle under the higher-derivative corrections. The higher-derivative correction to the Taub-NUT black hole was obtained in \cite{Chen:2024knw}. In this paper, we include the cosmological constant and construct the higher-derivative correction to the Taub-NUT-AdS black hole. We find that the local perturbative solution can be constructed straightforwardly. However, the analysis of its black hole thermodynamics becomes more challenging. The difficulty comes from multiple fronts. Although we continue to insist that the temperature and entropy are both derived using the standard and uncontroversial technique, the mass and the NUT charge become very difficult to determine directly, using the technique of \cite{Liu:2022wku}. This is because for a Killing vector $\xi$, the closed 2-form field associated with the Komar integration in Einstein gravity, namely
\be
\bf Q_\xi = {*d\xi}\,,
\ee
is no longer closed under the higher-derivative correction, i.e.~$d{\bf Q}_\xi\ne 0$. Of course, there must exist a generalized closed Komar 2-form ${\bf Q} + \delta {\bf Q}$; however, for general higher-derivative corrections, $\delta {\bf Q}$ is exceedingly difficult to evaluate. In many solutions, we can ignore the $\delta {\bf Q}$ contribution provided that we evaluate the Komar integration at the asymptotic infinity, as $\delta {\bf Q}$ decays faster than $\bf Q$. However, for Taub-NUT black holes, the evaluation requires the removing of the Misner string that extends from the horizon to the asymptotic infinity, so that the contribution from $\delta {\bf Q}$ cannot be ignored. Fortunately, as discussed, the bundle structure of the Taub-NUT-AdS black hole is unperturbed by the higher-derivative corrections. It follows from \cite{Liu:2022wku} that the NUT potential $\Phi_n=n/2$ is unperturbed and simple. Consider grand canonical ensemble with Gibbs free energy $G=M - T S - \Phi_n Q_n$, the first law is
\be
dG= -S dT - Q_n d\Phi_n\,,
\ee
where the quantities of $(S,T, \Phi_n)$ are known. We find that the consistency of the above equation allows us to determine $(G,Q_n)$ as well, up to an arbitrary function $g(n)$ of the NUT parameter.

To verify these thermodynamic quantities and determine the function $g(n)$, we adopt the Euclidean action approach. At first sight, this can be very difficult, as both the GHY $I_{\rm surf}$ and holographic counterterms $I_{\rm ct}$ can be very difficult to determine and known only for the Gauss-Bonnet or Lovelock combinations, e.g.~\cite{Liu:2008zf, Liu:2017kml}. Reall and Santos (RS) \cite{Reall:2019sah} showed, for the leading-order correction, that the correction to the Gibbs free energy could be simply obtained. The simplification has two folds: one is that it is unnecessary to construct the local perturbative solution; the other is that there is no contribution from the surface term for asymptotically-flat black holes. This technique was further generalized to the higher-order perturbations \cite{Ma:2023qqj}. When there is a cosmological constant, the situation becomes more complicated, but for the leading order correction, the perturbative solution remains unnecessary, but the surface and counterterms are still needed for the original RS method. This RS method was later enhanced and improved such that the boundary terms become also unnecessary \cite{Xiao:2023two,Hu:2023gru}. Both approaches have been successfully applied to various asymptotically AdS black holes. (See also, e.g.~\cite{Ma:2024ynp, Xiao:2025icr,Lu:2026clu}) We use both improved RS methods and show that the derived black hole thermodynamic quantities are correct and the function $g(n)$ is fully fixed.

The paper is organized as follows. In Section 2, we give a brief review of Taub-NUT-AdS black hole. In Section 3, we construct the leading-order perturbative
solution of the Taub-NUT-AdS black hole. In Section 4, we study the corresponding temperature, NUT potential and entropy corrections. Then apply the Maxwell relation to calculated remain quantities. In Section 5 we investigate the
field redefinition method proposed in \cite{Hu:2023gru} and demonstrate
its applicability to Taub-NUT-AdS spacetime. Using this method, all
thermodynamic quantities can be obtained consistently. Then, we also
apply the background reduction method proposed in \cite{Xiao:2023two} and
show that it leads to the same Gibbs free energy. Finally, we summarize
our results and discuss future directions in Section 6.

\section{Taub-NUT-AdS black hole in Einstein gravity}

In this section, we give a brief review of the Taub-NUT-AdS black hole in Einstein gravity and its thermodynamics. Einstein gravity coupled to a cosmological constant is described by the action
\begin{equation}
    I_{0}=\frac{1}{16\pi}\int d^4x\, \sqrt{-g}\, (R-2\Lambda) \,,
\end{equation}
where $R$ is the Ricci scalar. The cosmological constant is taken to be negative, given by $\Lambda=-3/l^2$ where $l$ is the resulting AdS radius. Einstein equation of motion is given by
\begin{equation}
	R_{ab}-\frac{1}{2}g_{ab}R-\frac{3}{l^2}g_{ab}=0\,.
\end{equation}
The Taub-NUT-AdS black hole provides an exact solution, given by
\begin{equation}
ds^2=-h(r)(dt+2n\cos\theta d\phi)^2+\frac{dr^2}{f(r)}+(r^2+n^2)(d\theta^2+\sin^2\theta d\phi^2),\label{nutansatz}
\end{equation}
where the metric functions are determined by the equation of motion as
\begin{equation}\label{Ein-NUT}
	f(r)=h(r)=\bar f(r)\equiv \frac{r^2-2mr-n^2}{r^2+n^2}-\frac{3n^4-6n^2r^2-r^4}{l^2(r^2+n^2)}.
\end{equation}
This metric admits two integration constants: $m$ and $n$. The position of the event horizon $r_0$ is determined from the condition $f(r_{0})=0$. The corresponding entropy and temperature can be computed directly
\begin{equation}
    T=\frac{\sqrt{f'(r_{0})h'(r_{0})}}{4\pi}=\frac{1}{4\pi r_{0}}(1+\frac{3(r_{0}^2+n^2)}{l^2})\,,\qquad S=\pi(r_{0}^2+n^2).
\end{equation}
When $n=0$, the solution reduces to the Schwarzschild-AdS black hole, and the Ricci-flat limit corresponds to $l\rightarrow\infty$. It is important to note that unlike the Schwarzschild black hole, even (arbitrary) negative values of $m$ can give rise to a positive real horizon $r_0$ in the Taub-NUT-AdS black hole.

There exist several formulations of thermodynamics for Taub-NUT-AdS black holes. In this work, we choose the thermodynamic scheme proposed in \cite{Liu:2023uqf}, which generalizes the discussion of the Ricci-flat Taub-NUT solution in \cite{Liu:2022wku} to include a cosmological constant. The degenerate Killing vectors of the Taub-NUT-AdS metric are located at the north and south poles
\begin{equation}
    l_{\pm}=\partial_\phi\mp 2n\partial_{t}=\partial_\phi\mp 4\Phi_{n}\partial_{t}\,.
\end{equation}
Motivated by the horizon Killing vector of Kerr black holes, $\partial_t + \Omega \partial_\phi$, where the angular velocity $\Omega$ is the thermodynamic potential conjugate to angular momentum, we set the NUT potential proportional to $n$ and adopt the expression $\Phi_{n}=n/2$. The corresponding conjugate NUT charge and mass are \cite{Liu:2023uqf}
\begin{equation}
    Q_{n}=\frac{n}{r_{0}}(1+\frac{3(n^2-r_{0}^2)}{l^2})\,,\qquad
    M=m+\frac{n^2}{r_{0}}(1+\frac{3(n^2-r_{0}^2)}{l^2})=m+2\Phi_{n}Q_{n}\,.
\end{equation}
Note that this definition of mass is always positive. Accordingly, these thermodynamic quantities obey the first law:
 \begin{equation}
     \delta M=T\delta S+\Phi_{n}\delta Q_{n}.
 \end{equation}
Furthermore, the Gibbs free energy can be obtained using the aforementioned thermodynamic quantities using a Legendre transformation. The identical result can also be calculated directly from the Euclidean action, given by
 \begin{equation}
     G=M-TS-\Phi_{n}Q_{n} =\frac{I}{\beta}\,.
 \end{equation}

\section{Taub-NUT-AdS black hole in higher-curvature corrections}

We now consider Einstein gravity extended by higher-curvature corrections. At first sight, the leading-order correction consists of a linear combination of quadratic curvature invariants: $c_{1} R_{abcd}R^{abcd}+c_{2}R_{ab}R^{ab}+c_{3}R^2$. However, for a perturbative theory, one may employ a field redefinition $g_{ab}\rightarrow g_{ab}+\lambda_{0}g_{ab}+\lambda_{1}g_{ab}R+\lambda_{2}R_{ab}$ to recast this general quadratic sector into the Gauss-Bonnet combination $R^2-4R_{ab}R^{ab}+R_{abcd}R^{abcd}$, which is a total derivative in four dimensions. Consequently, all quadratic curvature terms yield no nontrivial on-shell contribution in four dimensions and can be fully neglected. The true leading-order higher-curvature corrections therefore arise from the cubic curvature invariants, which can be expressed as
\begin{equation}
\begin{split}
\mathcal{L}_{\rm cubic}=\alpha_{1}R^3+\alpha_{2}RR_{ab}R^{ab}+\alpha_{3}R_{a}^{~b}R_{b}^{~c}R_{c}^{~a}
+\alpha_{4}R^{ab}R^{cd}R_{acbd} +\alpha_{5}RR_{abcd}R^{abcd}\\
+\alpha_{6}R^{ab}R_{acde}R_{b}^{~cde}+\alpha_{7}R^{ab}_{~~cd}R^{cd}_{~~ef}R^{ef}_{~~ab}
+\alpha_{8}R_{a~~b}^{~c~~d}R_{c~~d}^{~e~~f}R_{e~~f}^{~a~~b}.
\end{split}
\end{equation}
Note that some curvature polynomials are not all independent to each other, for example
 \begin{equation}
 {\rm Riem}^3_{\alpha_{1}}=4 {\rm Riem}^3_{\alpha_{2}}=16 {\rm Riem}^3_{\alpha_{3}}=16 {\rm Riem}^3_{\alpha_{4}}\,.
 \end{equation}
But in order not to lose generality, we shall retain all the combinations of Riemann curvature in this paper. The equations of motion (EOM) of Einstein gravity extended with the above cubic terms can be conveniently written as
\begin{equation}
    P_{acde}R_{b}^{~cde}-\frac{1}{2}g_{ab}L-2\nabla^{c}\nabla^{d}P_{acdb}=0
\end{equation}
where
\begin{equation}\label{Pabcd}
	P_{abcd}=\frac{\partial L}{\partial R^{abcd}}=P_{abcd}^{\rm E}+\Sigma_{i}\alpha_{i}  P_{abcd}^{i} \,,
\end{equation}
the components of tensor $P_{abcd}$ are
\begin{equation}
	\begin{split}
		&P_{abcd}^{\rm E}=\frac{1}{2}(g_{ac}g_{bd}-g_{ad}g_{bc}),\qquad P_{abcd}^{1}=
\frac{3}{2}(g_{ac}g_{bd}-g_{ad}g_{bc})R^2,
		\\&P_{abcd}^{2}=\frac{1}{2}(g_{ac}g_{bd}-g_{ad}g_{bc})R_{ef}R^{ef}+
\frac{1}{2}R(g_{bd}R_{ac}-g_{bc}R_{ad}-g_{ad}R_{bc}+g_{ac}R_{bd}),
		\\&P_{abcd}^{3}=\frac{3}{4}(g_{ac}R_{be}R_{d}^{e}-g_{ad}R_{be}R_{c}^{e}
-g_{bc}R_{ae}R_{d}^{e}+g_{bd}R_{ae}R_{c}^{e}),
		\\&P_{abcd}^{4}=\frac{1}{2}(g_{bd}R_{aecf}-g_{bc}R_{aedf}
-g_{ad}R_{becf}+g_{ac}R_{aedf})R^{ef}+\frac{1}{2}(R_{ac}R_{bd}-R_{ad}R_{bc}),
		\\&P_{abcd}^{5}=\frac{1}{2}(g_{ac}g_{bd}-g_{ad}g_{bc})R_{efgh}R^{efgh}
+2RR_{abcd},
		\\&P_{abcd}^{6}=\frac{1}{2}(R_{b}^{e}R_{aecd}+R_{d}^{e}R_{abce}-R_{c}^{e}R_{abde}
-R_{a}^{e}R_{becd})+\frac{1}{4}(g_{bd}R_{a}^{~efg}R_{cefg}
		\\&\qquad\qquad-g_{ad}R_{b}^{~efg}R_{cefg}+g_{ac}R_{b}^{~efg}R_{defg}-g_{bc}R_{a}^{~efg}R_{defg}),
		\\&P_{abcd}^{7}=3R_{ab}^{~~ef}R_{cdef},\qquad P_{abcd}^{8}=\frac{3}{2}(R_{a~~c}^{~e~~f}R_{bedf}-R_{a~~d}^{~e~~f}R_{becf}).
	\end{split}
\end{equation}
With the full EOM, we are in the position to construct the higher-order correction to the Taub-NUT-AdS black hole in Einstein gravity reviewed in the previous section. We adopt the same ansatz \eqref{nutansatz} whose time bundle preserves the spherical symmetry, but the metric functions $h(r)$ and $f(r)$ to be determined. Substituting this ansatz into the EOM yields a set of consistent equations of motion for $h$ and $f$. The full expressions are too lengthy to present here. We solve these equations perturbatively, working to the linear order in the coupling constants $\alpha_i$. The procedure for solving the EOM, together with the choice of integration constants, involves subtleties that are discussed in detail in Appendix~A. After a careful analysis, the first-order perturbative solution for the Taub-NUT-AdS black hole takes the following form:
 \begin{equation}\label{sol}
     \begin{split}
         &f(r)=\bar f+f_{0}+\frac{f_{1}}{l^2}+\frac{f_{2}}{l^4}+\frac{f_{3}}{l^6}\,,\\
         &h(r)=\bar f+h_{0}+\frac{h_{1}}{l^2}+\frac{h_{2}}{l^4}+\frac{h_{3}}{l^6}\,.
     \end{split}
 \end{equation}
 Here $\bar f$ denotes the leading Einstein solution given in \eqref{Ein-NUT}. The remaining parts of the metric functions correspond to the first-order corrections, which we organize as an expansion in powers of $l^{-1}$. Specifically, $f_0$ and $h_0$ are the first-order perturbative solutions for the Ricci-flat Taub--NUT spacetime, while $f_i$ and $h_i$ (with $i \neq 0$) encode the contributions arising from the AdS radius $l$. The explicit expressions for these functions are too complicated to present here; they are relegated to Appendix~B. At the flat limit $l\rightarrow \infty$, the metric reduces to the solution constructed in \cite{Chen:2024knw}.

We now examine the asymptotic behavior of the solution at large r. To this end, we series expand the metric function in inverse powers of $r$
\begin{equation}
\begin{split}
    \lim_{r\rightarrow\infty}f(r)=1+\frac{5n^2+r^2}{l^2}(1+\frac{9(16\alpha_{1}+4\alpha_{2}+\alpha_{3}
    +\alpha_{4})+6(4\alpha_{5}+\alpha_{6})+2(2\alpha_{7}+\alpha_{8})}{l^4})\,.
\end{split}
\end{equation}
The metric function $h(r)$ has the same asymptotic behavior at infinity because we have set it as one of the boundary conditions in the process of solving the EOM. The ultraviolet (UV) expansion shows that the effective AdS radius for our model with general cubic Riemann tensor couplings is related to the primary AdS radius via
\begin{equation}\label{leff}
l_{\rm eff}=l-\frac{9(16\alpha_{1}+4\alpha_{2}+\alpha_{3}+\alpha_{4})
+6(4\alpha_{5}+\alpha_{6})+2(2\alpha_{7}+\alpha_{8})}{2l^3}.
\end{equation}
\section{Black Hole Thermodynamics}

Based on the perturbative solution we obtained, we can now go a step further to study their contributions to the thermodynamic quantities. For the perturbative solution, we hold the mass parameter fixed rather than the horizon radius. To discuss the near-horizon thermodynamic quantities, we first identify the horizon location. The correction to the horizon can be directly obtained via $f(r_{h})=0=h(r_{h})$, namely
 \begin{equation}
     r_{h}=r_{0}+\Delta r_{h},
 \end{equation}
where
\begin{equation}
    \begin{split}
    \Delta r_{h}=&\frac{9(16\alpha_{1}+4\alpha_{2}+\alpha_{3}+\alpha_{4})r_0(3n^4-6n^2r_0^2-r_0^4)}{l^4(n^2+r_0^2)(l^2+3(n^2+r_0^2))}
        \\
    &+\frac{12\alpha_{5}+3\alpha_{6}}{2(l^4r_0^3(n^2+r_0^2)(l^2+3(n^2+r_0^2)))}[27n^8+78n^6r_0^2+84n^4r_0^4-6n^2r_0^6-7r_0^8
        \\ &+l^6(n^2-r_0^2)+l^2(27n^2-7r_0^2)(n^2+r_0^2)^2+l^4(9n^4+4n^2r_0^2-5r_0^4)]
        \\&+\frac{2\alpha_{7}+\alpha_{8}}{14l^4r_0^3(n^2+r_0^2)(l^2+3(n^2+r_0^2))}[l^6(27n^2-35r_0^2)+27l^4(9n^4+2n^2r_0^2-7r_0^4)
    \\&+3l^2(243n^6+387n^4r_0^2+37n^2r_0^4-91r_0^6)+3(243n^8+684n^6r_0^2+606n^4r_0^4
    \\&+84n^2r_0^6-49r_0^8)]
    \\&-\frac{9\alpha_{8}(l^4(n^2-r_0^2)+(9n^2- r_0^2)(n^2+r_0^2)^2+l^2(6 n^4 + 4n^2 r_0^2 - 2 r_0^4))}{4l^4r_0^3(n^2+r_0^2)}.
    \end{split}
\end{equation}
The Hawking temperature can be directly calculated via the standard method
\begin{equation}
    T=\frac{\sqrt{f'(r_{h})h'(r_{h})}}{4\pi}=T_{0}+\Delta T,
\end{equation}
where
\begin{equation}
    \begin{split}
        \Delta T=&[9(16\alpha_{1}+4\alpha_{2}+\alpha_{3}+\alpha_{4})+24\alpha_{5}+6\alpha_{6}]\frac{r_0(2l^2(3n^2+r_0^2)+3(9n^4+2n^2r_0^2+r_0^4))}{2l^6\pi(n^2+r_0^2)(l^2+3(n^2+r_0^2))}
        \\&-\frac{(2\alpha_{7}+\alpha_{8})}{28 l^6 \pi r_0^3 (n^2 + r_0^2)^2 (l^2 + 3 (n^2 + r_0^2))}[l^8 (5 n^2 - 7 r_0^2) + 6 l^6 (10 n^4 + 3 n^2 r_0^2 - 7 r_0^4)
 \\&+6l^4(45n^6 +79n^4 r_0^2 + 27n^2 r_0^4 -7r_0^6)+6l^2(90n^8+267n^6r_0^2+273n^4r_0^4 +89 n^2 r_0^6\\&-7r_{0}^8)
 +9(45n^{10}+177n^8 r_0^2 + 222 n^6 r_0^4 + 118 n^4 r_0^6 + 21 n^2 r_0^8 -
  7 r_0^{10})].
    \end{split}
\end{equation}
We employ the Wald entropy formula \cite{Wald:1993nt,Iyer:1994ys} to compute the entropy for the perturbative Taub-NUT-AdS black hole solution
\begin{equation}
     S_{\rm Wald}=\frac{1}{8\pi}\int d^2x\sqrt{h} \epsilon^{ab}\epsilon^{cd}P_{abcd}\,,
\end{equation}
where $h$ is the determinant of the induced metric on the horizon, $\epsilon_{ab}$ is the binormal to the horizon satisfying $\epsilon_{ab}\epsilon^{ab}=-2$, and $P_{abcd}$ is defined in (\ref{Pabcd}). In Einstein gravity, this is equal to one-quarter of the area of the event horizon. However, in higher-derivative gravity, corrections from higher-order terms arise. Specifically, the Wald entropy of our linearly perturbed solution is
\begin{equation}
    S_{\rm Wald}=\pi(r_0^2+n^2)+\Delta S \,,
\end{equation}
with
\begin{equation}
    \begin{split}
        \Delta S=&9\pi(16\alpha_{1}+4\alpha_{2}+\alpha_{3}+\alpha_{4})[\frac{3(r_{0}^2+n^2)}{l^4}-\frac{2r_{0}^2(-3n^4+6n^2r_{0}^2+r_{0}^4)}{l^4(r_{0}^2+n^2)(l^2+3(n^2+r_{0}^2))}]
        \\&+6(4\alpha_{5}+\alpha_{6})\frac{\pi[3n^6+15n^4r_0^2-3n^2r_0^4+r_0^6-2l^4(n^2+r_0^2)-5l^2(n^2+r_0^2)^2]}{l^4(n^2+r_0^2)(l^2+3(n^2+r_0^2))}
        \\&-\frac{2\pi(2\alpha_{7}+\alpha_{8})}{14l^4r_0^2(n^2+r_0^2)(l^2+3(n^2+r_0^2))}[l^6(15n^2-7r_0^2)+9l^4(15n^4+22n^2r_0^2+7r_0^4)\\&+3l^2(135n^6+369n^4r_0^2+341n^2r_0^4+91r_0^6)+3(135n^8+450n^6r_0^2+528n^4r_0^4\\&+294n^2r_0^6+49r_0^8)]+\frac{18\alpha_{8}\pi(l^2+n^2+r_0^2)}{l^4}.
    \end{split}
\end{equation}
The degenerate Killing vectors at the north and south poles remain intact under the perturbation. It is similar to the Ricci-flat case in\cite{Chen:2024knw}. Hence, the NUT potential is natural fixed. Specifically, for the modified Taub-NUT-AdS black hole solution, the degenerate Killing vectors at the north and south poles remain unchanged under perturbation,
\begin{equation}
    \xi=\partial_{\phi}\pm 2n\partial_{t}
\end{equation}
and so does the NUT potential
\begin{equation}
    \Phi_{n}=\Phi_{n0}=\frac{n}{2}.
\end{equation}
The form of the black hole first law is as expected, which should be:
\begin{equation}
	\delta M = T \delta S + \Phi_n \delta Q_n.
\end{equation}
There are a total of five thermodynamic quantities for the black hole. Currently, we can directly obtain three of them, namely the temperature, entropy, and NUT potential. The remaining two quantities, the black hole mass $M$ and NUT charge $Q_n$ can be obtained by the Maxwell relation. To do so, we find that it is more convenient to first perform a Legendre transformation of the black hole mass
\begin{equation}
    G=M-TS-\Phi_{n}\delta Q_{n},
\end{equation}
where $G$ is the Gibbs free energy. The first law of the black hole thermodynamics can then be written as
\begin{equation}
    \delta G=-S\delta T-Q_{n}\delta \Phi_{n}.
\end{equation}
In this grand canonical ensemble, the thermodynamic variables are $T$ and $\Phi_n=n/2$.
Therefore, the NUT charge and the entropy satisfy the following Maxwell relation:
\begin{equation}\label{Maxwell}
    (\frac{\partial S}{\partial \Phi_{n}})\Big|_{T}=(\frac{\partial Q_{n}}{\partial T})\Big|_{\Phi_{n}}.
\end{equation}
Equivalently, the NUT charge can be obtained directly as
\begin{equation}
    Q_{n}=2[\int \frac{\partial}{\partial n}(S\frac{\partial T}{\partial r_{0}})dr_{0}-S\frac{\partial T}{\partial n}].
\end{equation}
There are two independent parameters $(r_{0},n)$ in our perturbative Taub-NUT-AdS solution, which implies that we should select two thermodynamic quantities as variables. For convenience, we first take the unperturbed temperature and NUT potential as the variables. Correspondingly, all other parameters will change accordingly. We find that under the redefinition of parameters
\begin{equation}
    r_{0}\rightarrow \tilde{r}_{0}+\delta r_{0}~,\qquad n\rightarrow\tilde{n}\,,
\end{equation}
where
\begin{equation}
    \begin{split}
        \delta r_{0}=&[9(16\alpha_{1}+4\alpha_{2}+\alpha_{3}+\alpha_{4})+6(4\alpha_{5}+\alpha_{6})]\frac{3\tilde{r}_{0}^3 (2 l^2 (3 \tilde{n}^2 + \tilde{r}_{0}^2) + 3 (9 \tilde{n}^4 + 2 \tilde{n}^2 \tilde{r}_{0}^2 + \tilde{r}_{0}^4))}{l^4 (l^2 + 3 \tilde{n}^2 - 3 \tilde{r}_{0}^2) (\tilde{n}^2 + \tilde{r}_{0}^2) (l^2 + 3 (\tilde{n}^2 + \tilde{r}_{0}^2))}
        \\&+\frac{-(2\alpha_{7}+\alpha_{8})}{7 l^4 \tilde{r}_{0} (\tilde{n}^2 + \tilde{r}_{0}^2)^2 (l^4 + 6 l^2 \tilde{n}^2 + 9 \tilde{n}^4 - 9 \tilde{r}_{0}^4)}[l^8 (5 \tilde{n}^2 - 7 \tilde{r}_{0}^2)+ 6 l^6 (10 \tilde{n}^4 + 3 \tilde{n}^2 \tilde{r}_{0}^2 - 7 \tilde{r}_{0}^4)
        \\&+ 6 l^4 (45 \tilde{n}^6 + 79 \tilde{n}^4 \tilde{r}_{0}^2+27 \tilde{n}^2 \tilde{r}_{0}^4 - 7 \tilde{r}_{0}^6)+6 l^2 (90 \tilde{n}^8 + 267 \tilde{n}^6 \tilde{r}_{0}^2+273 \tilde{n}^4 \tilde{r}_{0}^4 + 89 \tilde{n}^2 \tilde{r}_{0}^6
        \\&- 7 \tilde{r}_{0}^8) +9 (45 \tilde{n}^{10} + 177 \tilde{n}^8 \tilde{r}_{0}^2 + 222 \tilde{n}^6 \tilde{r}_{0}^4 + 118 \tilde{n}^4 \tilde{r}_{0}^6 +21 \tilde{n}^2 \tilde{r}_{0}^8 - 7 \tilde{r}_{0}^{10})].
    \end{split}
\end{equation}
The temperature and NUT potential are fixed at the linear order of $\alpha$
\begin{equation}
    T=T_{0}=\frac{1}{4\pi\tilde{r}_{0}}(1+\frac{3(\tilde{r}_{0}^2+\tilde{n}^2)}{l^2})+\mathcal{O}(\alpha_{i}^2)~,~~\Phi_{n}=\Phi_{n0}=\frac{\tilde{n}}{2}+\mathcal{O}(\alpha_{i}^2).
\end{equation}
The Wald entropy turns out to be
\begin{equation}\label{S-fixT}
    \begin{split}
        &\tilde{S}_{\rm Wald}=\pi(\tilde{r}_{0}^2+\tilde{n}^2)+\Delta \tilde{S}
        \\&\Delta \tilde{S}=\frac{27(16\alpha_{1}+4\alpha_{2}+\alpha_{3}+\alpha_{4})(l^2 + 3 \tilde{n}^2 - \tilde{r}_{0}^2) (\tilde{n}^2 + \tilde{r}_{0}^2)}{l^4 (l^2 + 3 \tilde{n}^2 - 3 \tilde{r}_{0}^2)}
        \\&-\frac{(4\alpha_{5}+\alpha_{6})(2 l^4 + l^2 (5 \tilde{n}^2 - 7 \tilde{r}_{0}^2) - 3 (\tilde{n}^2 + \tilde{r}_{0}^2)^2)}{l^4 (l^2 + 3 \tilde{n}^2 - 3 \tilde{r}_{0}^2)}
        \\&+\frac{3\pi(2\alpha_{7}+\alpha_{8})}{7 l^4 \tilde{r}_{0}^2 (l^2 + 3 \tilde{n}^2 - 3 \tilde{r}_{0}^2)(\tilde{n}^2 + \tilde{r}_{0}^2)^2}[3 l^4(\tilde{n}^2 + \tilde{r}_{0}^2)^2 (15 \tilde{n}^2 + 7 \tilde{r}_{0}^2) +
 9 (\tilde{n}^2 + \tilde{r}_{0}^2)^3
 \\&(15 \tilde{n}^4 - 7 \tilde{r}_{0}^4) +
 l^6 (5 \tilde{n}^4 + 6 \tilde{n}^2 \tilde{r}_{0}^2 - 7 \tilde{r}_{0}^4) +
 9 l^2 (\tilde{n}^2 + \tilde{r}_{0}^2)^2 (15 \tilde{n}^4 + 16 \tilde{n}^2 \tilde{r}_{0}^2 - 7 \tilde{r}_{0}^4)]
 \\&+\frac{28\pi\alpha_{8}\pi(l^2+\tilde{n}^2+\tilde{r}_{0}^2)}{l^4}\,.
    \end{split}
\end{equation}
Then by applying the Maxwell relation (\ref{Maxwell}), we can obtain the NUT charge
\begin{equation}
    Q_{n}=\frac{n}{r_{0}}(1+\frac{3(n^2-r_{0}^2)}{l^2})+\Delta Q_{n}+g(n),
\end{equation}
where $g(n)$ is an arbitrary function and
\begin{equation}
 \begin{split}
 \Delta Q_{n}=&\frac{27\tilde{n}(16\alpha_{1}+4\alpha_{2}+\alpha_{3}+\alpha_{4})  (l^4 + 9 \tilde{n}^4 - 30 \tilde{n}^2 \tilde{r}_{0}^2 + 9 \tilde{r}_{0}^4 +
   l^2 (6 \tilde{n}^2 - 8 \tilde{r}_{0}^2))}{l^6 \tilde{r}_{0} (l^2 + 3 \tilde{n}^2 - 3 \tilde{r}_{0}^2)}
   \\&+\frac{6\tilde{n} (4\alpha_{5}+\alpha_{6})(l^4 + 6 l^2 (\tilde{n}^2 - 2 \tilde{r}_{0}^2) + 9 (\tilde{n}^4 - 6 \tilde{n}^2 \tilde{r}_{0}^2 + \tilde{r}_{0}^4))}{l^6 \tilde{r}_{0} (l^2 + 3 \tilde{n}^2 - 3 \tilde{r}_{0}^2)}
   \\&+\frac{12\tilde{n}(2\alpha_{7}+\alpha_{8})[(-9 l^4 + 45 l^2 \tilde{r}_{0}^2 + 108 \tilde{n}^2 \tilde{r}_{0}^2)(\tilde{n}^2 + \tilde{r}_{0}^2)^2-6 l^6(\tilde{n}^2 +\tilde{r}_{0}^2)-l^8]}{7 l^6 \tilde{r}_{0} (l^2 + 3 \tilde{n}^2 - 3 \tilde{r}_{0}^2)(\tilde{n}^2 + \tilde{r}_{0}^2)^2}
   \\&+\frac{18 \tilde{n}\alpha_{8}  (l^2 + 3 \tilde{n}^2 - 3 \tilde{r}_{0}^2)}{l^6 \tilde{r}_{0}}.
 \end{split}
\end{equation}
Once the NUT charge and entropy satisfy the integrability condition, the Gibbs free energy can be obtained directly by integration,
\begin{equation}\label{GG}
    \begin{split}
        &G=\frac{-\sigma(3 \tilde{n}^4 + \tilde{r}_{0}^4 + l^2 (\tilde{n}^2 - \tilde{r}_{0}^2))}{4l^2 \tilde{r}_{0}}+\Delta G-\int \frac{g(n)}{2}dn,
        \\&\Delta G=\frac{-9 \sigma^3 (16\alpha_{1}+4\alpha_{2}+\alpha_{3}+\alpha_{4})(3 l^2 (\tilde{n}^2 - \tilde{r}_{0}^2) + (-3 \tilde{n}^2 + \tilde{r}_{0}^2)^2)}{4 l^6 \tilde{r}_{0}}
        \\&+\frac{3\sigma^3(4\alpha_{5}+\alpha_{6})(2 l^4 - 3 \tilde{n}^4 + 6 \tilde{n}^2 \tilde{r}_{0}^2 + \tilde{r}_{0}^4 + l^2 (5 \tilde{n}^2 + 7 \tilde{r}_{0}^2))}{2 l^6 \tilde{r}_{0}}
        \\&+\frac{\sigma^3(2\alpha_{7}+\alpha_{8})}{28 l^6 \tilde{r}_{0}^3 (\tilde{n}^2 + \tilde{r}_{0}^2)}[l^6 (5 \tilde{n}^2 - 7 \tilde{r}_{0}^2) + 9 l^4 (5 \tilde{n}^4 + 12 \tilde{n}^2 \tilde{r}_{0}^2 + 7 \tilde{r}_{0}^4) +27 l^2 (5 \tilde{n}^6
        \\&+ 21 \tilde{n}^4 \tilde{r}_{0}^2 + 23 \tilde{n}^2 \tilde{r}_{0}^4 + 7 \tilde{r}_{0}^6)+9(15 \tilde{n}^8 + 60 \tilde{n}^6 \tilde{r}_{0}^2 + 66 \tilde{n}^4 \tilde{r}_{0}^4 + 28 \tilde{n}^2 \tilde{r}_{0}^6 + 7 \tilde{r}_{0}^8)]
        \\&+\frac{36\tilde{n}\sigma^3\alpha_{8}(l^2+3\tilde{n}^2-3\tilde{r}_{0}^2) }{l^6 \tilde{r}_{0}}.
    \end{split}
\end{equation}
Correspondingly, the black hole mass can be recovered by performing the inverse Legendre transformation,
\begin{equation}
    M=G+TS_{W}+\Phi_{n}Q_{n}.
\end{equation}
Thus, we see that by knowing unambiguously the quantities $(S,T,\Phi_n)$, the integrability condition of the first law allows us to determine the remaining $(M,Q_n)$, up to only one arbitrary function $g(n)$ of the NUT parameter. In the next section, we adopt the Euclidean action approach and determine this function completely.

\section{Euclidean action approach via the RS method}

As discussed in the introduction, the RS approach to Euclidean action method renders the construction of the perturbative solution unnecessary, (of course) provided that such a black hole solution actually exists, which we established in the previous section. For the asymptotic AdS black holes, the contribution from the boundary terms to the Euclidean action cannot be generally ignored in the original proposal of the RS method. Here we considered two enhanced or improved versions that do not require the computation of the boundary terms.

\subsection{Method 1: field redefinition}

An improved version was proposed to avoid the computation of the complicated boundary terms. The core idea is to first employ the field redefinition to recast generic higher-derivative curvature terms into Weyl-invariant combinations, thereby eliminating the contribution from the boundary. One then reverses the field redefinition to obtain the general results.

\subsubsection{Results of the RS method in Weyl gravity}

We first directly apply the RS method in Weyl gravity, a special subclass of higher-derivative gravitational theories. All higher-order curvature terms in Weyl gravity consist solely of contractions of the Weyl tensor. For instance, at cubic order, the Lagrangian reads
\begin{equation}
\begin{split}
    &\mathcal{L}_{Weyl}=\alpha {\rm Weyl}+\beta \widetilde {\rm Weyl},
    \\&{\rm Weyl}=C^{ab}_{~~cd}C^{cd}_{~~ef}C^{ef}_{~~ab}\,,\qquad \widetilde {\rm Weyl}=C_{a~~b}^{~c~~d}C_{c~~d}^{~e~~f}C_{e~~f}^{~a~~b}.
\end{split}
\end{equation}
Only two such cubic Weyl tensor combinations exist, yet these two structures are linearly dependent
\begin{equation}
    {\rm Weyl}=2\widetilde {\rm Weyl} \,.
\end{equation}
We therefore only need to consider the full Lagrangian featuring a single cubic Weyl invariant
\begin{equation}
    \mathcal{L}=R-2\Lambda+\alpha C^{ab}_{~~cd}C^{cd}_{~~ef}C^{ef}_{~~ab}.
\end{equation}
Owing to the traceless property of the Weyl tensor, the theory admits no correction to the cosmological constant. Furthermore, the integral of the Weyl Lagrangian $\mathcal{L}_{\text{Weyl}}$ is free of divergences. All surface terms decay to zero in the large-$r$ limit, circumventing the technical obstacles encountered in holographic renormalization procedures. For Taub-NUT-AdS black hole backgrounds, we take the Hawking temperature and NUT potential as the fundamental thermodynamic variables. The Gibbs free energy corresponding to our thermodynamic ensemble reads
\begin{equation}
    G(T_{0},\Phi_{n_0})=T_{0}I_{\rm total}\,,
\end{equation}
where $I_{\rm total}$ denotes the full Euclidean action. To implement the RS formalism for our setup, we substitute the zeroth-order Taub-NUT-AdS solution into the gravitational action, using only the leading-order solution. We find
\begin{equation}
\begin{split}
   G_{\rm RS}=&-\frac{(3\tilde{n}^{4}+\tilde{r}_{0}^{4}+l^2(\tilde{n}^2-\tilde{r}_{0}^2))}{4l^2\tilde{r}_{0}}+\frac{\alpha}{28 l^6 \tilde{r}_{0}^3 (\tilde{n}^2 + \tilde{r}_{0}^2)}[135 \tilde{n}^8 + 414 \tilde{n}^6 \tilde{r}_{0}^2 + 384 \tilde{n}^4 \tilde{r}_{0}^4
    \\&+ 98 \tilde{n}^2 \tilde{r}_{0}^6 - 7 \tilde{r}_{0}^8 + l^6 (5 \tilde{n}^2 - 7 \tilde{r}_{0}^2)+3l^4 (15 \tilde{n}^4 + 8 \tilde{n}^2 \tilde{r}_{0}^2 - 7 \tilde{r}_{0}^4)+3l^2 (45 \tilde{n}^6 + 91 \tilde{n}^4 \tilde{r}_{0}^2
    \\&+39 \tilde{n}^2 \tilde{r}_{0}^4 -7\tilde{r}_{0}^6)].
\end{split}
\end{equation}

\subsubsection{Improved RS method for general cubic curvature gravity}

The general cubic gravity may be recast into Weyl gravity via suitable field redefinitions. While the field redefinition derived in \cite{Hu:2023gru} holds for arbitrary spacetime dimensions, our analysis restricts to the four-dimensional case  $D=4$. Accordingly, the corresponding transformation for our setup reads
\begin{equation}
\begin{split}
        g_{\mu\nu}\rightarrow g'_{\mu\nu}=&g_{\mu\nu}+\beta_{0}g_{\mu\nu}+\sigma^2\beta_{1}R_{\mu\alpha\beta\gamma}R_{\nu}^{~\alpha\beta\gamma}+\sigma^2\beta_{2}R^{\rho\sigma}R_{\mu\rho\nu\sigma}+\sigma^2\beta_{3}RR_{\mu\nu}
        \\&+\sigma^2\beta_{4}R_{\mu}^{~\rho}R_{\nu\rho}+\sigma^2g_{\mu\nu}(\beta_{5}R^2+\beta_{6}R_{\rho\sigma}R^{\rho\sigma}+\beta_{7}R_{\rho\sigma\alpha\beta}R^{\rho\sigma\alpha\beta})
        \\&+\frac{\sigma^2}{l^2}(\gamma_{1}R_{\mu\nu}+\gamma_{2}g_{\mu\nu}R).
\end{split}
\end{equation}
where
\begin{equation}
    \begin{split}
        &\beta_{0}=\frac{\sigma^2(1152\alpha_{1}+288\alpha_{2}+72\alpha_{4}+192\alpha_{5}+48\alpha_{6}+59\alpha_{7}+16\alpha_{8})}{4l^4},
        \\&\beta_{1}=\alpha_{6}+5\alpha_{7}-\frac{3\alpha_{8}}{2}~,~~\beta_{2}=\alpha_{4}-6\alpha_{7}+3\alpha_{8},
        \\&\beta_{4}=\alpha_{3}-6\alpha_{7}+2\alpha_{8}~,~~\beta_{5}=-\frac{\beta_{3}}{2}-\alpha_{1}+\frac{5\alpha_{7}}{9}-\frac{2\alpha_{8}}{9},
        \\&\beta_{6}=\beta_{3}-\alpha_{2}-\frac{\alpha_{3}+\alpha_{4}+\alpha_{8}}{2}+\alpha_{7}~,~~\beta_{7}=-\alpha_{5}-2\alpha_{7}-\frac{\alpha6+\alpha_{8}}{2},
        \\&\gamma_{1}=3(4\beta_{3}-4\alpha_{2}-\alpha_{3}-\alpha_{4}-8\alpha_{5}-2\alpha_{6}-12\alpha_{7}+4\alpha_{8}),
        \\&\gamma_{2}=-\beta_{3}+12\alpha_{1}+6\alpha_{2}+\frac{3(\alpha_{3}+\alpha_{4})}{2}+8\alpha_{5}+2\alpha_{6}+\frac{28\alpha_{7}}{3}-\frac{17\alpha_{8}}{6}.
    \end{split}
\end{equation}
The field redefinition may contribute a Weyl combination of 4-derivative level. Under the field redefinition, our theory can be transformed to
\begin{equation}
\begin{split}
       &I=\frac{1}{16\pi}\int\sqrt{-g}(\mathcal{L}_{\rm EG}+\mathcal{L}_{\rm hd1}+\mathcal{L}_{\rm hd2})\,,
       \\&\mathcal{L}_{\rm EG}=\sigma'(R+\frac{6}{l'^2})~,~~\mathcal{L}_{\rm hd1}=q'\sigma'^2C_{\mu\nu\rho\sigma}C^{\mu\nu\rho\sigma}\,,
\\&\mathcal{L}_{\rm hd2}=\sigma'^3(\alpha_{7}'C^{ab}_{~~cd}C^{cd}_{~~ef}C^{ef}_{~~ab}
+\alpha_{8}'C_{a~~b}^{~c~~d}C_{c~~d}^{~e~~f}C_{e~~f}^{~a~~b}) \,,
\end{split}\label{reducedweylaction}
\end{equation}
where the redefined coupling constants are
\begin{equation}\label{ReDePar}
	\begin{split}  &\sigma'=\sigma+\frac{3\sigma^3(576\alpha_{1}+144\alpha_{2}+12\alpha_{3}+36\alpha_{4}+96\alpha_{5}+24\alpha_{6}+25\alpha_{7}+8\alpha_{8})}{4l^4},
		\\&l'=l-\frac{\sigma^2(576\alpha_{1}+144\alpha_{2}-36\alpha_{3}+36\alpha_{4}+96\alpha_{5}+24\alpha_{6}+43\alpha_{7}+8\alpha_{8})}{8l^3},
		\\&q'=-\frac{3(8\alpha_{5}+2\alpha_{6}+4\alpha_{7}-\alpha_{8})\sigma}{2l^2}~,\qquad
\alpha_{7}'=\alpha_{7}~,\qquad\alpha_{8}'=\alpha_{8}\,.
	\end{split}
\end{equation}
The higher-derivative bulk terms now involve only the Weyl invariants, and hence the on-shell action is convergent for the unperturbed Taub-NUT-AdS black hole. Substituting the solution, we find
\begin{equation}
\begin{split}
&I_{\rm EG}=-\frac{\sigma'(3\hat{n}^{4}+\hat{r}_{0}^{4}+l'^2(\hat{n}^2-\hat{r}_{0}^2))}{4l'^2\hat{r}_{0}},
    \qquad I_{\rm hd1}=-\frac{\sigma'^2 (l'^4 + 3 \hat{n}^4 + \hat{r}_{0}^4 + 2 l'^2 (2 \hat{n}^2 + \hat{r}_{0}^2))}{l'^4 \hat{r}_{0}},
\\&I_{\rm hd2}=\frac{\sigma'^3(2\alpha_{7}'+\alpha_{8}')}{28 l'^6 \hat{r}_{0}^3 (\hat{n}^2 + \hat{r}_{0}^2)}[135 \hat{n}^8 + 414 \hat{n}^6 \hat{r}_{0}^2 + 384 \hat{n}^4 \hat{r}_{0}^4 + 98 \hat{n}^2 \hat{r}_{0}^6 - 7 \hat{r}_{0}^8 + l'^6 (5 \hat{n}^2 - 7 \hat{r}_{0}^2)
 \\&+3l'^4 (15 \hat{n}^4 + 8 \hat{n}^2 \hat{r}_{0}^2 - 7 \hat{r}_{0}^4)+3l'^2 (45 \hat{n}^6 + 91 \hat{n}^4 \hat{r}_{0}^2+39 \hat{n}^2 \hat{r}_{0}^4 -7\hat{r}_{0}^6)]\,.
\end{split}
\end{equation}
As stated earlier, the Hawking temperature and NUT potential in the RS method are fixed to be
\begin{equation}
    \hat{T}=\frac{1}{4\pi\hat{r}_{0}}(1+\frac{3(\hat{r}_{0}^2+\hat{n}^2)}{l'^2})\,,\qquad \hat{\Phi}_{n}=\frac{\hat{n}}{2}.
\end{equation}
According to (\ref{ReDePar}), the effective AdS radius $\ell'$ depends on the coupling constants of the general theory before the field redefinition. It follows that the temperature will also depend on coefficients of higher derivative terms. Hence we need to redefine the integration constants ($\hat{r}_{0},\hat{n}$) to ($\tilde{r}_{0},\tilde{n}$) that make them independent of $\alpha_{i}$.
\begin{equation}
\hat{r}_{0}\rightarrow\tilde{r}_{0}+\delta\hat{r}_{0}~,\qquad \hat{n}\rightarrow\tilde{n}.
\end{equation}
where
\begin{equation}
    \begin{split}
        \delta\hat{r}_{0}=&-\frac{[27(16\alpha_{1}+4\alpha_{2}+\alpha_{3}+\alpha_{4})+18(4\alpha_{5}+\alpha_{6})+6(2\alpha_{7}+\alpha_{8})]\sigma^2 \tilde{r}_{0} (\tilde{n}^2 + \tilde{r}_{0}^2)}{l^4 (l^2 + 3 \tilde{n}^2 - 3 \tilde{r}_{0}^2)}
        \\&+\frac{3 \tilde{r}_{0} (\tilde{n}^2 + \tilde{r}_{0}^2)\beta_{0}}{l^2+3\tilde{n}^2 - 3 \tilde{r}_{0}^2}.
    \end{split}
\end{equation}
After this redefinition, the temperature and NUT potential are fixed to be the same as the unperturbed Taub-NUT-AdS solution, namely
\begin{equation}
    \tilde{T}=\frac{1}{4\pi\tilde{r}_{0}}(1+\frac{3(\tilde{r}_{0}^2+\tilde{n}^2)}{l'^2})~,
    \qquad \tilde{\Phi}_{n}=\frac{\tilde{n}}{2}.
\end{equation}
The general perturbed Gibbs free energy, with temperature and NUT potential kept fixed, can be readily calculated, given by
\begin{equation}
    \begin{split}
        &G_{\rm RS}=\frac{-\sigma(3 \tilde{n}^4 + \tilde{r}_{0}^4 + l^2 (\tilde{n}^2 - \tilde{r}_{0}^2))}{4l^2 \tilde{r}_{0}}+\Delta G_{\rm RS},
        \\&\Delta G_{\rm RS}=\frac{-9 \sigma^3 (16\alpha_{1}+4\alpha_{2}+\alpha_{3}+\alpha_{4})(3 l^2 (\tilde{n}^2 - \tilde{r}_{0}^2) + (-3 \tilde{n}^2 + \tilde{r}_{0}^2)^2)}{4 l^6 \tilde{r}_{0}}
        \\&+\frac{3\sigma^3(4\alpha_{5}+\alpha_{6})(2 l^4 - 3 \tilde{n}^4 + 6 \tilde{n}^2 \tilde{r}_{0}^2 + \tilde{r}_{0}^4 + l^2 (5 \tilde{n}^2 + 7 \tilde{r}_{0}^2))}{2 l^6 \tilde{r}_{0}}
        \\&+\frac{\sigma^3(2\alpha_{7}+\alpha_{8})}{28 l^6 \tilde{r}_{0}^3 (\tilde{n}^2 + \tilde{r}_{0}^2)}[l^6 (5 \tilde{n}^2 - 7 \tilde{r}_{0}^2) + 9 l^4 (5 \tilde{n}^4 + 12 \tilde{n}^2 \tilde{r}_{0}^2 + 7 \tilde{r}_{0}^4) +27 l^2 (5 \tilde{n}^6
        \\&+ 21 \tilde{n}^4 \tilde{r}_{0}^2 + 23 \tilde{n}^2 \tilde{r}_{0}^4 + 7 \tilde{r}_{0}^6)+9(15 \tilde{n}^8 + 60 \tilde{n}^6 \tilde{r}_{0}^2 + 66 \tilde{n}^4 \tilde{r}_{0}^4 + 28 \tilde{n}^2 \tilde{r}_{0}^6 + 7 \tilde{r}_{0}^8)]
        \\&+\frac{36\tilde{n}\sigma^3\alpha_{8}(l^2+3\tilde{n}^2-3\tilde{r}_{0}^2) }{l^6 \tilde{r}_{0}}.
    \end{split}
\end{equation}
Remarkably, this result obtained using the RS method not only is consistent with our earlier approach by constructing the perturbative solution, but also fixes the previous undetermined function $g(n)$, namely $g(n)=0$.

\subsection{Method 2: background subtraction}

An alternative scheme for resolving the divergence of the Euclidean action, without appealing to the boundary terms, is the background subtraction formalism \cite{Hawking:1982dh}. Its application to RS method was introduced in \cite{Xiao:2023two}. The Euclidean action obtained by evaluating the renormalized integral of the action, given by
\begin{equation}
    \int_{\mathcal{M}}\rightarrow \int_{\mathcal{M}}^{reg}=\lim_{r_{c}\rightarrow\infty}\bigg(\int^{\rm (BH)}_{\mathcal{M}}-\sqrt{\frac{g_{tt}^{\rm (BH)}(r_{c})}{g_{tt}^{\rm (AdS)}(r_{c})}}\int^{\rm (AdS)}_{\mathcal{M}}\bigg).
\end{equation}
This method likewise bypasses the complications arising from evaluating boundary terms and obviates the need for holographic renormalization. This construction has already been successfully implemented within the Wald formalism for Einstein gravity, where it reproduces the exact first law of thermodynamics for Kerr-AdS black holes \cite{Xiao:2023lap}. In what follows, we adopt this formalism to derive the full thermodynamic relations for Taub-NUT-AdS black holes.

\subsubsection{A brief review of background Reduction in Einstein Gravity}
We first briefly recapitulate this method by examining the thermodynamics of Taub-NUT-AdS black holes within Einstein gravity. Under this prescription, the Euclidean action takes the form
\begin{equation}\label{XY-Ein}
    I=\frac{1}{16\pi}\int_{\mathcal{M}}^{\rm reg}(R-2\Lambda).
\end{equation}
The Euclidean action was evaluated for the Schwarzschild-AdS black holes in \cite{Hawking:1982dh}. We applied the method for Taub-NUT-AdS black hole, for which the integrand simplifies to  $R-2\Lambda=2\Lambda$. The resulting integral over the black hole manifold is given by
\begin{equation}
    \frac{1}{16\pi}\int^{\rm (BH)}_{\mathcal{M}}(R-2\Lambda)=\frac{\Lambda}{2}\int_{r_{c}}^{r_{0}}(r^2+n^2)dr
    =-\frac{\Lambda(r_c^3-r_0^3-3n^2(r_0-r_{c}))}{6}.
\end{equation}
The AdS vacuum of the regular integration is (the upper limit of integration should be 0)
\begin{equation}
   \frac{1}{16\pi}\int^{\rm (AdS)}_{\mathcal{M}}(R-2\Lambda)=\frac{\Lambda}{2}\int_{r_{c}}^{0}(r^2+n^2)dr
   =-\frac{\Lambda(3n^2r_{c}+r_{c}^3)}{6}.
\end{equation}
Then the $tt$- component of both black hole and AdS vacuum are
\begin{equation}
g_{tt}^{\rm (BH)}=\frac{-3n^4\Lambda+3n^2(1+2r^2\Lambda)+r(6m-3r+r^3\Lambda)}{3(n^2+r^2)}\,,\qquad g_{tt}^{\rm (AdS)}=-1+\frac{(5n^2+r^2)\Lambda}{3}\,.
\end{equation}
Substituting all expressions into Eq. (\ref{XY-Ein}) and subsequently taking the UV limit, we arrive at
\begin{equation}
    I=\frac{3(r_0^2-n^2+n^4\Lambda)+r_{0}^4\Lambda}{12r_{0}} \,.
\end{equation}
This result matches the outcome derived via conventional renormalization, where Gibbons- Hawking boundary terms and counterterms are introduced.

\subsubsection{Background subtraction in higher derivative gravity}

We now turn to  higher derivative gravity. Adopting the procedure outlined above, we only need to compute
\begin{equation}
    I=\frac{1}{16\pi}\int_{\mathcal{M}}^{\rm reg}(R-2\Lambda_{e}+\mathcal{L}_{hd}+2(\Lambda_{e}-\Lambda))
\end{equation}
where $\Lambda_{e}$ is effective cosmological constant that corresponds to cosmological constant of 0-order solution. Tracing the Einstein equation, we can obtain
\begin{equation}
    \Lambda_{e}-\Lambda= \sum_{k}\frac{k-D}{2D}\mathcal{L}_{\rm hd}^{\rm AdS}
\end{equation}
where $k$ counts the number of derivatives acting on the metric. In our case, it is
\begin{equation}
    \Lambda=\Lambda_{e}-(16\alpha_{1}+4\alpha_{2}+\alpha_{3}+\alpha_{4}+\frac{2(4\alpha_{5}+\alpha_{6})}{3}+\frac{2(2\alpha_{7}+\alpha_{8})}{9})\Lambda_{e}^3.
\end{equation}
This result agrees with the UV expansion of our perturbative solution in Eq. (\ref{leff}). We further treat $\bar{T}_{0}$ and $\bar{\Phi}_{n0}$ as our fundamental thermodynamic variables. Nevertheless, it is critical to stress that the fixed temperature reads
\begin{equation}
\bar{T}_{0}=\frac{1}{4\pi\bar{r}_{0}}(1+\frac{3(\bar{r}_{0}^2+\bar{n}^2)}{l_{\rm eff}^2})\,,
\end{equation}
where $l_{\rm eff}$ is the effective AdS radius. Thus we use ($\bar{r}_{0},\bar{n}$) in this section instead of ($\tilde{r}_{0},\tilde{n}$) to emphasize the difference between them. And the on-shell action can be evaluated using the unperturbed metric, which gives
 \begin{equation}
 \begin{split}
     &\frac{1}{16\pi}\int_{\mathcal{M}}^{\rm reg}(\mathcal{L}_{\rm hd})
     \\&=\beta[\frac{(16\alpha_{1}+4\alpha_{2}+\alpha_{3}+\alpha_{4})\Lambda_{e}^2 (3(\bar{r}_{0}^2-\bar{n}^2+\bar{n}^4\Lambda_{e})+\bar{r}_{0}^4 \Lambda_{e})}{6\bar{r}_{0}}+\frac{(4\alpha_{5}+\alpha_{6}) \Lambda_{e} (-1+(\bar{n}^2+\bar{r}_{0}^2)\Lambda_{e})}{\bar{r}_{0}}]
     \\&-\frac{\beta\alpha_{7}}{378\bar{r}_{0}^3(\bar{r}_{0}^2+\bar{n}^2)}[135\bar{n}^8\Lambda_{e}^3+3\bar{n}^6\Lambda_{e}^2(-135+194\bar{r}_{0}^2\Lambda_{e})+3\bar{n}^4\Lambda_{e}(135-581\bar{r}_{0}^2\Lambda_{e}+184\bar{r}_{0}^4\Lambda_{e}^2)    \\&+7\bar{r}_{0}^2(27+81\bar{r}_{0}^2\Lambda_{e}-75\bar{r}_{0}^4\Lambda_{e}^2+7\bar{r}_{0}^6\Lambda_{e}^3)+\bar{n}^2(-135+972\bar{r}_{0}^2\Lambda_{e}-1863\bar{r}_{0}^4\Lambda_{e}^2+154\bar{r}_{0}^6\Lambda_{e}^3)]
     \\&-\frac{\beta\alpha_{8}}{756 \bar{r}_{0}^3 (\bar{n}^2 + \bar{r}_{0}^2)}[135 \bar{n}^8 \Lambda_{e}^3 + 3\bar{n}^6 \Lambda_{e}^2 (-135 + 68 \bar{r}_{0}^2 \Lambda_{e})+3\bar{n}^4\Lambda_{e}(135-77\bar{r}_{0}^2 \Lambda_{e} + 58\bar{r}_{0}^4 \Lambda_{e}^2)
    \\&+7\bar{r}_{0}^2 (27 - 81 \bar{r}_{0}^2 \Lambda_{e} +
    33 \bar{r}_{0}^4 \Lambda_{e}^2 - 11 \bar{r}_{0}^6 \Lambda_{e}^3) +
 \bar{n}^2 (-135 - 162 \bar{r}_{0}^2 \Lambda_{e} +
    405 \bar{r}_{0}^4 \Lambda_{e}^2 + 28 \bar{r}_{0}^6 \Lambda_{e}^3)],
 \end{split}
 \end{equation}
 \begin{equation}
 \begin{split}
      \frac{1}{16\pi}\int_{\mathcal{M}}^{\rm reg}2(\Lambda_{e}-\Lambda)=&(16\alpha_{1}+4\alpha_{2}+\alpha_{3}+\alpha_{4}+\frac{2(4\alpha_{5}+\alpha_{6})}{3}+\frac{2(2\alpha_{7}+\alpha_{8})}{9})
      \\&\times\frac{\beta\Lambda_{e}^2(-3\bar{n}^2+3\bar{r}_{0}^2+3\bar{n}^4\Lambda_{e}+\bar{r}_{0}^4\Lambda_{e})}{12\bar{r}_{0}}.
 \end{split}
 \end{equation}
The Gibbs' free energy can be written as $\bar{G}=\bar{T}_{0}I$. We then shall return to the framework of ($\tilde{r}_{0},\tilde{n}$). We find that under the parameter redefinition
\begin{equation}
    \bar{r}_{0}\rightarrow \tilde{r}_{0}+\delta\bar{r}_{0},\qquad \bar{n}\rightarrow\tilde{n},
\end{equation}
where
\begin{equation}
\begin{split}           \delta\bar{r}_{0}=\frac{[27(16\alpha_{1}+4\alpha_{2}+\alpha_{3}+\alpha_{4})
+6(6\alpha_{5}+3\alpha_{6}+2\alpha_{7}+\alpha_{8})]
\bar{r}_{0}(\bar{r}_{0}^2+\bar{n}^2)}{l^4(l^2+3(\bar{n}^2-\bar{r}_{0}^2))}\,.
\end{split}
\end{equation}
The temperature and NUT potential are fixed as
\begin{equation}
    \bar{T}_{0}(\bar{r}_{0},\bar{n},l_{\rm eff})\rightarrow T_{0}(\tilde{r}_{0},\tilde{n},l)~,\qquad \bar{\Phi}_{n0}(\bar{n})\rightarrow\Phi_{n0}(\tilde{n}).
\end{equation}
Upon performing the parameter redefinition, the Gibbs free energy derived via the background reduction technique coincides with the expression obtained by recasting the theory into Weyl gravity through field redefinitions, satisfying $\bar{G}=G$. Accordingly, the undetermined function $g(n)$ also vanishes.

\section{Conclusion}

In this work, we investigated cubic curvature corrections to both the four-dimensional Taub-NUT-AdS black hole solution and the corresponding black hole thermodynamics. Working to the first order in the higher-derivative couplings, we obtained the complete analytic corrections to the metric functions. The solution captures both the shift in the effective AdS radius and the corrections to the horizon geometry, with the asymptotically locally flat result recovered in the vanishing cosmological constant limit.

Using the perturbative solution, we calculated the corrected Hawking temperature and Wald entropy. The NUT potential remains unchanged, reflecting the preservation of the fibre bundle of the Taub-NUT-AdS geometry. Taking the temperature and NUT potential as independent thermodynamic variables, we reconstructed the NUT charge and Gibbs free energy through the Maxwell relation. We found that thermodynamic integrability can almost determine the mass and the NUT charge, up to only an arbitrary function $g(n)$ of the NUT parameter $n$.

We then adopted the Euclidean action approach to resolve this ambiguity by applying two improved versions of the RS method. The corrections to Gibbs free energy with fixed temperature and NUT potential were computed from the bulk action of cubic curvature terms evaluated on the original unperturbed Taub-NUT-AdS black hole. We found that the result is fully consistent with the earlier method based on explicit perturbative solution and furthermore it sets the function $g(n)=0$. We therefore established the thermodynamics of Taub-NUT-AdS black hole under the general cubic-curvature tensor corrections. The black hole thermodynamics satisfies both the Wald entropy formula and the first law.

Our results demonstrate two main points. One is that the formalism of the Taub-NUT-AdS black hole thermodynamics proposed in \cite{Liu:2022wku,Liu:2023uqf} can withstand the check of high-derivative corrections. In particular, the NUT potential is simply given by $\Phi_n=n/2$, independent of higher-derivative corrections. The other is that the improved RS prescriptions can be consistently applied to the asymptotically locally AdS geometries considered here, despite the additional structure introduced by the NUT parameter and Misner strings. The analytic perturbative solution provides a direct benchmark for evaluating higher-derivative thermodynamics, while the agreement between the two action-based approaches supplies an independent determination beyond integrability alone. It would be interesting to extend this analysis to charged and rotating NUT spacetimes, higher perturbative orders, and thermodynamic ensembles associated with different global treatments of the Misner strings.

\section*{Acknowledgment}

We are grateful to Peng-Ju Hu and Liang Ma for the useful discussion.  This work is supported in part by NSFC (National Natural Science Foundation of China) Grant No.~12575061 and  No.~12375052, Tianjin University Graduate Liberal Arts and Sciences Innovation Award Program (2023) No. B1 2023-005 and Tianjin University Self-Innovation Fund Extreme Basic Research Project Grant No. 2025XJ21-0007.

\section*{Appendix}
\appendix
\section{Details of Solving Tricks and Boundary Condition}

The equations of motion of higher derivative gravity can be written as
\begin{equation}
    P_{acde}R_{b}^{~cde}-\frac{1}{2}g_{ab}L-2\nabla^{c}\nabla^{d}P_{acdb}=0
\end{equation}
where the $P_{abcd}$ has been given in Section 3.

To construct the linear perturbation solution, we consider the following expansion
\begin{equation}
\begin{split}
   &h(r)=h_{EG}+\Sigma_{i}\alpha_{i}\Delta h_{i}(r)
   \\&f(r)=f_{EG}+\Sigma_{i}\alpha_{i}\Delta f_{i}(r)
\end{split}
\end{equation}
with
 \begin{equation}
     f_{EG}=h_{EG}=\frac{r^2-2mr-n^2}{r^2+n^2}-\frac{3n^4-6n^2r^2-r^4}{l^2(r^2+n^2)}.
 \end{equation}
We submit the metric into the EOM and extract the linear differential equations of $ \Delta f_{i}$ and $\Delta h_{i}$ at the linear $\alpha_{i}$ order. The $rr$ component of EOM yields $\Delta f_{i}(r)$ as a function of $\Delta h_{i}(r)$ and $\Delta h_{i}'(r)$, namely $\Delta f_{i}=\mathcal{F}(\Delta h_{i},\Delta h_{i}')$. Take this formula back to the EOM, the $rr$ component is eliminated and the $tt$ component gives a second order differential equation of $\Delta h_{i}$. The equations of function $\Delta h_{i}$ are mutually independent. Solving the second order equations of $\Delta h_{i}$, we find the corrected part of the metric functions can be simply concluded as
 \begin{equation}\label{antz}
     \Delta f_{i}=\sum_{j=0}^{3}\frac{f_{i}^{(j)}}{l^{2j}}+f^{i}_{bdy}
 \end{equation}
where $f_{bdy}^{i}$ is the boundary term that is associate with the integral constant$(c_{1}^{i},c_{2}^{i})$. We find
\newpage
\begin{equation}
    \begin{split}
        &f_{1}^{(0)}=0,~h_{1}^{(0)}=0,~f_{2}^{(0)}=0,~h_{2}^{(0)}=0,~f_{3}^{(0)}=0,~h_{3}^{(0)}=0,~f_{4}^{(0)}=0,~h_{4}^{(0)}=0,
        \\&f_{5}^{(0)}=\frac{48}{r^2(r^2+n^2)^7}[m^3(3n^8r+38n^6r^3-360n^4r^5+234n^2r^7-11r^9)+m(-3n^{10}r\\&-226n^8r^3+900n^6r^5-702n^4r^7+95n^2r^9)+m^2(n^{10}+12n^8r^2-564n^6r^4+1054n^4r^6\\&-285n^2r^8+6r^{10})-n^2(n^{10}+24n^8r^2-192n^6r^4+290n^4r^6-129n^2r^8+6r^{10})]\\&h_{5}^{(0)}=\frac{48}{r(r^2+n^2)^7}[n^2r(3n^8-46n^6r^2+60n^4r^4-18n^2r^6+r^8)+m^3(n^8-18n^6r^2\\&+60n^4r^4-46n^2r^6+3r^8)-mn^2(n^8-54n^6r^2+192n^4r^4-122n^2r^6+15r^8)\\&-m^2(15n^8r-122n^6r^3+192n^4r^5-54n^2r^7+r^9)]
        \\&f_{6}^{(0)}=\frac{12}{r^2(r^2+n^2)^7}[m^3(3n^8r+38n^6r^3-360n^4r^5+234n^2r^7-11r^9)+m(-3n^{10}r\\&-226n^8r^3+900n^6r^5-702n^4r^7+95n^2r^9)+m^2(n^{10}+12n^8r^2-564n^6r^4+1054n^4r^6\\&-285n^2r^8+6r^{10})-n^2(n^{10}+24n^8r^2-192n^6r^4+290n^4r^6-129n^2r^8+6r^{10})]\\&h_{6}^{(0)}=\frac{12}{r(r^2+n^2)^7}[n^2r(3n^8-46n^6r^2+60n^4r^4-18n^2r^6+r^8)+m^3(n^8-18n^6r^2\\&+60n^4r^4-46n^2r^6+3r^8)-mn^2(n^8-54n^6r^2+192n^4r^4-122n^2r^6+15r^8)\\&-m^2(15n^8r-122n^6r^3+192n^4r^5-54n^2r^7+r^9)]\\&f_{7}^{(0)}=\frac{8}{7r^2(r^2+n^2)^7}[-3mn^2r(27n^8+2116n^6r^2-8598n^4r^4+6900 n^2r^6-973 r^8)\\&+m^3(81n^8r+1036n^6r^3-10290n^4r^5+6972n^2r^7-343r^9)+3m^2(9 n^{10}+111n^8r^2\\&-5298n^6r^4+10226n^4r^6-2871n^2r^8+63r^{10})-n^2(27n^{10}+669n^8r^2-5478n^6r^4\\&+8390n^4r^6-3861n^2r^8+189r^{10})]
       \\&h_{7}^{(0)}=\frac{8}{7r(r^2+n^2)^7}\Big[4n^4r(15n^6-237n^4r^2+265n^2r^4-27r^6)
-12m^2(33n^8r\cr
&-219n^6r^3+247n^4r^5-45n^2r^7)-3mn^2(9n^8-404n^6r^2+1230n^4r^4\cr
&-516n^2r^6+17r^8)+m^3(27n^8-476n^6r^2+1050n^4r^4-588n^2r^6+35r^8)\Big] \\&f_{8}^{(0)}=\frac{2}{7r^2(r^2+n^2)^7}[n^4(9n^8+174n^6r^2-1140n^4r^4+1490n^2r^6-405r^8)\\&+3mn^2r(9n^8+514n^6r^2-1704n^4r^4+942n^2r^6-49r^8)+m^2(-9n^{10}-90n^8r^2\\&+3744n^6r^4-5046n^4r^6+729n^2r^8)+m^3(-27n^8r -322n^6r^3+2100n^4r^5\\&-798n^2r^7+7r^9)]\\&h_{8}^{(0)}=\frac{2}{7r(r^2+n^2)^7}[n^2r(-69n^8+1002n^6r^2-1660n^4r^4+918n^2r^6-63r^8)\\&+m^3(-9n^8+182n^6r^2-1680n^4r^4+1722n^2r^6-119r^8)+m(9n^{10}-978n^8r^2\\&+4716n^6r^4-4590n^4r^6+843n^2r^8)+3m^2(51n^8r-810n^6r^3+2056n^4r^5\\&-774n^2r^7+21r^9)]
    \end{split}
\end{equation}
\begin{equation}
    \begin{split}       &f_{1}^{(1)}=0,~h_{1}^{(1)}=0,~f_{2}^{(1)}=0,~h_{2}^{(1)}=0,~f_{3}^{(1)}=0,~h_{3}^{(1)}=0,~f_{4}^{(1)}=0,~h_{4}^{(1)}=0
    \\&f_{5}^{(1)}=\frac{48}{r^2 (n^2 + r^2)^7}[-11 n^{14} - 258 n^{12} r^2 + 2151 n^{10} r^4 - 3732 n^8 r^6 +1611 n^6 r^8+ 54 n^4 r^{10}\\&-7 n^2 r^{12}-24 m n^2 r (n^{10} + 66 n^8 r^2 - 300 n^6 r^4 + 258 n^4 r^6 -21 n^2 r^8- 4 r^{10})+m^2 (3 n^{12}\\&+ 18 n^{10} r^2 - 2103 n^8 r^4 + 4468 n^6 r^6 -1203 n^4 r^8 - 102 n^2 r^{10} + 7 r^{12})],
    \\&h_{5}^{(1)}=\frac{-192}{r (n^2 + r^2)^7}[m^2 n^2 r (15 n^8 - 122 n^6 r^2 + 192 n^4 r^4 - 54 n^2 r^6 +r^8) - 3 n^4 r (3 n^8\\&-46n^6r^2+60 n^4 r^4 -18 n^2 r^6 + r^8) +2 m n^4 (n^8 - 54 n^6 r^2 + 192 n^4 r^4 - 122 n^2 r^6 + 15 r^8)],
    \\&f_{6}^{(1)}=\frac{12}{r^2 (n^2 + r^2)^7}[-11 n^{14} - 258 n^{12} r^2 + 2151 n^{10} r^4 - 3732 n^8 r^6 +1611 n^6 r^8+ 54 n^4 r^{10}\\&-7 n^2 r^{12}-24 m n^2 r (n^{10} + 66 n^8 r^2 - 300 n^6 r^4 + 258 n^4 r^6 -21 n^2 r^8- 4 r^{10})+m^2 (3 n^{12}\\&+ 18 n^{10} r^2 - 2103 n^8 r^4 + 4468 n^6 r^6 -1203 n^4 r^8 - 102 n^2 r^{10} + 7 r^{12})],
    \\&h_{6}^{(1)}=\frac{-48}{r (n^2 + r^2)^7}[m^2 n^2 r (15 n^8 - 122 n^6 r^2 + 192 n^4 r^4 - 54 n^2 r^6 +r^8) - 3 n^4 r (3 n^8\\&-46n^6r^2+60 n^4 r^4 -18 n^2 r^6 + r^8) +2 m n^4 (n^8 - 54 n^6 r^2 + 192 n^4 r^4 - 122 n^2 r^6 + 15 r^8)],
    \\&f_{7}^{(1)}=\frac{4}{7r^2(n^2+r^2)^7}(162m^2n^{12}-594n^{14}-1296mn^{12}r+1008m^2n^{10}r^2-14400n^{12}r^2
      \\&-88992mn^{10}r^3-118278m^2n^8r^4+122598n^{10}r^4+411552mn^8r^5+259440m^2n^6r^6
      \\&-215376n^8r^6-365184mn^6r^7-73386m^2n^4r^8+97146n^6r^8+32496mn^4r^9
      \\&-6048m^2n^2r^{10}+3024n^4r^{10}+6048mn^2r^{11}+462m^2r^{12}-462n^2r^{12})
 \\&h_{7}^{(1)}=-\frac{48}{7r(n^2+r^2)^7}[-123n^{12}r+1947n^{10}r^3-2086n^8r^5+126n^6r^7-63n^4r^9+7n^2r^{11}\\&+4mn^2(9n^{10}-414n^8r^2+1254n^6r^4-456n^4r^6+25n^2r^8-18r^{10})+m^2(267n^{10}r\\&-1803n^8r^3+1942n^6r^5-270n^4r^7+63n^2r^9-7r^{11})]\\&f_{8}^{(1)}=\frac{-6}{7 r^2 (n^2 + r^2)^7}[-33 n^{14} - 618 n^{12} r^2 + 4305 n^{10} r^4 - 6580 n^8 r^6 +1449 n^6 r^8\\&+ 126 n^4 r^{10} + 7 n^2 r^{12} -8 m n^4 r (9 n^8 + 450 n^6 r^2 - 1752 n^4 r^4 + 1038 n^2 r^6 +31 r^8)\\&+m^2 (9 n^{12} + 42 n^{10} r^2 - 4737 n^8 r^4 + 7348 n^6 r^6 -801 n^4 r^8 - 126 n^2 r^{10} - 7 r^{12})]\\&h_{8}^{(1)}=\frac{24}{7 r (n^2 + r^2)^7}[-66 n^{12} r + 951 n^{10} r^3 - 1694 n^8 r^5 + 1008 n^6 r^7 -7 n^2 r^{11} +m (6 n^{12} \\&- 612 n^{10} r^2 + 3048 n^8 r^4 - 3300 n^6 r^6 +530 n^4 r^8 + 72 n^2 r^{10}) +m^2 (48 n^{10} r - 759 n^8 r^3\\&+ 2090 n^6 r^5 - 864 n^4 r^7 -42 n^2 r^9 + 7 r^{11})]
    \end{split}
\end{equation}
\begin{equation}
    \begin{split}
        &f_{1}^{(2)}=\frac{-144(n^4+3n^2(m-r)r+2mr^3)}{r^2 (n^2 + r^2)},~h_{1}^{(2)}=\frac{-144 (mn^2-3n^2r+r^3)}{r(n^2 + r^2)}
        \\&f_{2}^{(2)}=\frac{-36(n^4+3n^2(m-r)r+2mr^3)}{r^2 (n^2 + r^2)},~h_{2}^{(2)}=\frac{-144 (mn^2-3n^2r+r^3)}{r(n^2 + r^2)}
        \\&f_{3}^{(2)}=\frac{-9(n^4+3n^2(m-r)r+2mr^3)}{r^2 (n^2 + r^2)},~h_{3}^{(2)}=\frac{-36(mn^2-3n^2r+r^3)}{r(n^2 + r^2)}
        \\&f_{4}^{(2)}=\frac{-9(n^4+3n^2(m-r)r+2mr^3)}{r^2 (n^2 + r^2)},~h_{4}^{(2)}=\frac{-36(mn^2-3n^2r+r^3)}{r(n^2 + r^2)}
        \\&f_{5}^{(2)}=\frac{-24}{r^2 (n^2 + r^2)^7}[81 n^{16} + 3 n^{12} (1820 m - 5329 r) r^3 +9 n^8(3018 m - 1493 r) r^7\\&+ 3 n^2 (5 m - r) r^{13} + 2 m r^{15} +5 n^4 r^{11} (-144 m + 19 r) + 9 n^{14} r (11 m + 203 r)\\&-n^6 r^9 (907 m + 1479 r) + n^{10} r^5 (-28743 m + 31847 r)]\\&h_{5}^{(2)}=\frac{-24}{r (n^2 + r^2)^7}[-291 n^{14}r+ 4399n^{12}r^3 -5799 n^{10}r^5 + 1683 n^8 r^7 -121 n^6 r^9\\&- 3 n^4 r^{11} + 3 n^2 r^{13} + r^{15} +m n^2 (33 n^{12} - 1722 n^{10} r^2 + 6159 n^8 r^4 - 3884 n^6 r^6\\&+495 n^4 r^8 + 6 n^2 r^{10} + r^{12})]\\&f_{6}^{(2)}=\frac{-4}{r^2 (n^2 + r^2)^7}[81 n^{16} + 3 n^{12} (1820 m - 5329 r) r^3 +9 n^8(3018 m - 1493 r) r^7\\&+ 3 n^2 (5 m - r) r^{13} + 2 m r^{15} +5 n^4 r^{11} (-144 m + 19 r) + 9 n^{14} r (11 m + 203 r)\\&-n^6 r^9 (907 m + 1479 r) + n^{10} r^5 (-28743 m + 31847 r)]\\&h_{6}^{(2)}=\frac{-4}{r (n^2 + r^2)^7}[-291 n^{14}r+ 4399n^{12}r^3 -5799 n^{10}r^5 + 1683 n^8 r^7 -121 n^6 r^9\\&- 3 n^4 r^{11} + 3 n^2 r^{13} + r^{15} +m n^2 (33 n^{12} - 1722 n^{10} r^2 + 6159 n^8 r^4 - 3884 n^6 r^6\\&+495 n^4 r^8 + 6 n^2 r^{10} + r^{12})]
        \\&f_{7}^{(2)}=\frac{4}{7r^2(n^2+r^2)^7}(-2167n^{16}-2613mn^{14}r-50997n^{14}r^2-152958mn^{12}r^3
        \\&+454917n^{12}r^4+820485mn^{10} r^5-917393n^{10}r^6-798756mn^8r^7+406827n^8r^8\\&+36261mn^6r^9+42609n^6r^{10}+24066mn^4r^{11}-3577n^4r^{12}-21mn^2r^{13}+21n^2r^{14})
        \\&h_{7}^{(2)}=-\frac{4}{7r(n^2+r^2)^7}[-6069n^{14}r+95785n^{12}r^3-98769n^{10}r^5+1413n^8 r^7-6223n^6r^9
    \\&+651n^4r^{11}+21n^2r^{13}+7r^{15}+mn^2(871n^{12}-40662n^{10}r^2+122793 n^8r^4
    -37876n^6r^6\\&+3273n^4r^8-3414n^2r^{10}+7r^{12})]
    \\&f_{8}^{(2)}=\frac{2}{7 r^2 (n^2 + r^2)^7}[353 n^{16}+ n^{12} (18388 m - 48579 r) r^3 +9 n^8 (5978 m - 1669 r) r^7 \\&+ n^6 (4733 m - 2751 r) r^9 -21 n^2 (5 m - r) r^{13} - 14 m r^{15} - 7 n^4 r^{11} (48 m + 7 r)\\&+3 n^{14} r (137 m + 2153 r) + n^{10} r^5 (-86799 m + 86575 r)]\\&h_{8}^{(2)}=\frac{2}{7 r (n^2 + r^2)^7}[-3003 n^{14} r + 43319 n^{12} r^3 - 82671 n^{10} r^5 + 53019 n^8 r^7 +3199 n^6 r^9\\&- 651 n^4 r^{11} - 21 n^2 r^{13} - 7 r^{15} +m n^2 (137 n^{12} - 13770 n^{10} r^2 + 70743 n^8 r^4 -85100 n^6 r^6\\&+ 11847 n^4 r^8 + 3414 n^2 r^{10} - 7 r^{12})]
    \end{split}
\end{equation}
\begin{equation}
    \begin{split}
        &f_{1}^{(3)}=\frac{144(-3n^6+15n^4r^2+3n^2r^4+r^6)}{r^2 (n^2 + r^2)}
        \\&h_{1}^{(3)}=\frac{576 (3n^4 -n^2r^2)}{(n^2 + r^2)}
        \\&f_{2}^{(3)}=\frac{36(-3n^6+15n^4r^2+3n^2r^4+r^6)}{r^2 (n^2 + r^2)}
        \\&h_{2}^{(3)}=\frac{144 (3n^4 -n^2r^2)}{(n^2 + r^2)}
        \\&f_{3}^{(3)}=\frac{9(-3n^6+15n^4r^2+3n^2r^4+r^6)}{r^2 (n^2 + r^2)}
        \\&h_{3}^{(3)}=\frac{36(3n^4 -n^2r^2)}{(n^2 +r^2)}
        \\&f_{4}^{(3)}=\frac{9(-3n^6+15n^4r^2+3n^2r^4+r^6)}{r^2 (n^2 + r^2)}
        \\&h_{4}^{(3)}=\frac{36(3n^4 -n^2r^2)}{(n^2 + r^2)}
        \\&f_{5}^{(3)}=\frac{24}{r^2 (n^2 + r^2)^7}[-99 n^{18} - 2115 n^{16} r^2 + 19728 n^{14} r^4 - 45000 n^{12} r^6 +18834 n^{10} r^8 \\&+ 3546 n^8 r^{10} - 72 n^6 r^{12} + 48 n^4 r^{14} +9 n^2 r^{16} + r^{18}]\\&h_{5}^{(3)}=\frac{96 n^2 }{(n^2 + r^2)^7}(99 n^{14} - 1455 n^{12} r^2 + 1959 n^{10} r^4 - 531 n^8 r^6 +57 n^6 r^8 + 3 n^4 r^{10}\\&- 3 n^2 r^{12} - r^{14})\\&f_{6}^{(3)}=\frac{4}{r^2 (n^2 + r^2)^7}[-99 n^{18} - 2115 n^{16} r^2 + 19728 n^{14} r^4 - 45000 n^{12} r^6 +18834 n^{10} r^8 \\&+ 3546 n^8 r^{10} - 72 n^6 r^{12} + 48 n^4 r^{14} +9 n^2 r^{16} + r^{18}]\\&h_{6}^{(3)}=\frac{24n^2}{(n^2 + r^2)^7}(99 n^{14} - 1455 n^{12} r^2 + 1959 n^{10} r^4 - 531 n^8 r^6 +57 n^6 r^8 + 3 n^4 r^{10}\\&- 3 n^2 r^{12} - r^{14})\\&f_{7}^{(3)}=\frac{4}{7r^2(n^2+r^2)^7}(-2613n^{18}-59157n^{16}r^2+559536n^{14}r^4- 1296888n^{12}r^6\\&+568062n^{10}r^8+98742n^8r^{10}-6328n^6r^{12}+336n^4r^{14}+63n^2r^{16}+7r^{18})\\&h_{7}^{(3)}=\frac{16n^2}{7 (n^2 + r^2)^7}(2085n^{14}-32665n^{12}r^2+32561n^{10}r^4+1179n^8r^6+3199n^6r^8\\&-315n^4r^{10}-21n^2r^{12}-7r^{14})
        \\&f_{8}^{(3)}=\frac{2}{7 r^2 (n^2 + r^2)^7}(411 n^{18} + 7371 n^{16} r^2 - 60384 n^{14} r^4 + 126408 n^{12} r^6\\&-15570 n^{10} r^8 - 4074 n^8 r^{10} + 728 n^6 r^{12} + 336 n^4 r^{14} +63 n^2 r^{16} + 7 r^{18})\\&h_{8}^{(3)}=\frac{-8n^2}{7 (n^2 + r^2)^7}(939 n^{14} - 13703 n^{12} r^2 + 27919 n^{10} r^4 - 19323 n^8 r^6\\&-2191 n^6 r^8 + 315 n^4 r^{10} + 21 n^2 r^{12} + 7 r^{14})
    \end{split}
\end{equation}
\begin{equation}
    \begin{split}
        &f^{i}_{bdy}=c_{1}^{(i)}[\frac{6n^6-5r^3+6n^2(-5+r)r^3+3n^4r^2(-12+5r)}{(-3-10n^2-15n^4+l^2(-1+(-3+2m)n^2))r^2(n^2+r^2)}
        \\&+\frac{l^2(2n^4+(-3+4m)r^3+3n^2r(2m+(-2+r)r))}{(-3-10n^2-15n^4+l^2(-1+(-3+2m)n^2))r^2(n^2+r^2)}]
        \\&+c_{2}^{(i)}[\frac{-3n^6+r^3+n^2(10-3r)r^3-3n^4r^2(-6+5r)}{(-3-10n^2-15n^4+l^2(-1+(-3+2m)n^2))r^2(n^2+r^2)}
        \\&+\frac{l^2(-n^4+(1-2m)r^3+n^2 r(-3(-1 + r)r+m(-3+r^2)))}{(-3-10n^2-15n^4+l^2(-1+(-3+2m)n^2))r^2(n^2+r^2)}]
        \\&h^{i}_{bdy}=c_{1}^{(i)}[\frac{r(15n^4(-2+r)+10n^2r(-3+2r)+r(-5+2r^3))}{(-3-10n^2- 15n^4+l^2(-1+(-3+2m)n^2))r(n^2+r^2)}
        \\&+\frac{l^2(2mn^2+r(3n^2(-2+r)+r(-3+2r)))}{(-3-10n^2- 15n^4+l^2(-1+(-3+2m)n^2))r(n^2+r^2)}]
        \\&+c_{2}^{(i)}[\frac{-(-1 + r) (r(15n^4+r+10n^2 r+r^2+r^3)}{(-3-10n^2-15n^4+l^2(-1+(-3+2m)n^2))r(n^2+r^2)}
        \\&+\frac{-(-1 + r)l^2 (-mn^2(1+r)+r(3n^2+r))}{(-3-10n^2-15n^4+l^2(-1+(-3+2m)n^2))r(n^2+r^2)}]
    \end{split}
\end{equation}

To eliminate the dimensionally anomalous terms in the boundary contributions, the values of the two class of parameters $(c_{1}^{(i)},c_{2}^{(i)})$ must be fixed by imposing appropriate boundary conditions.
The $(c_{1}^{(i)},c_{2}^{(i)})$ are decided as follow:

First, we require that there must be the same asymptotic behaviors between $\Delta h_{i}$ and $\Delta f_{i}$, which are related to the effective cosmological constants.
 We find
 \begin{equation}
     \begin{split}
         &c_{2}^{(1)}=\frac{144(1-(-3 + 2m)n^2)}{l^4}+\frac{144(3+10n^2+15n^4)}{l^6}+2c_{1}^{(1)}
         \\&c_{2}^{(2)}=\frac{36(1-(-3 + 2m)n^2)}{l^4}+\frac{36(3+10n^2+15n^4)}{l^6}+2c_{1}^{(2)}
         \\&c_{2}^{(3)}=\frac{9(1-(-3 + 2m)n^2)}{l^4}+\frac{9(3+10n^2+15n^4)}{l^6}+2c_{1}^{(3)}
         \\&c_{2}^{(4)}=\frac{9(1-(-3 + 2m)n^2)}{l^4}+\frac{9(3+10n^2+15n^4)}{l^6}+2c_{1}^{(4)}
         \\&c_{2}^{(5)}=\frac{24(1-(-3 + 2m)n^2)}{l^4}+\frac{24(3+10n^2+15n^4)}{l^6}+2c_{1}^{(5)}
         \\&c_{2}^{(6)}=\frac{6(1-(-3 + 2m)n^2)}{l^4}+\frac{6(3+10n^2+15n^4)}{l^6}+2c_{1}^{(6)}
         \\&c_{2}^{(7)}=\frac{4(1-(-3 + 2m)n^2)}{l^4}+\frac{4(3+10n^2+15n^4)}{l^6}+2c_{1}^{(7)}
         \\&c_{2}^{(8)}=\frac{2(1-(-3 + 2m)n^2)}{l^4}+\frac{2(3+10n^2+15n^4)}{l^6}+2c_{1}^{(8)}
     \end{split}
 \end{equation}

 Furthermore, the terms with anomalous dimension must vanish. It gives
 \begin{equation}
     \begin{split}
         &c_{1}^{(1)}=\frac{144(1+10n^2-15n^4+l^2(1+(-3+m)n^2))}{l^6}
         \\&c_{1}^{(2)}=\frac{36(1+10n^2-15n^4+l^2(1+(-3+m)n^2))}{l^6}
         \\&c_{1}^{(3)}=\frac{9(1+10n^2-15n^4+l^2(1+(-3+m)n^2))}{l^6}
         \\&c_{1}^{(4)}=\frac{9(1+10n^2-15n^4+l^2(1+(-3+m)n^2))}{l^6}
         \\&c_{1}^{(5)}=\frac{24(1+10n^2-15n^4+l^2(1+(-3+m)n^2))}{l^6}
         \\&c_{1}^{(6)}=\frac{6(1+10n^2-15n^4+l^2(1+(-3+m)n^2))}{l^6}
         \\&c_{1}^{(7)}=\frac{4(1+10n^2-15n^4+l^2(1+(-3+m)n^2))}{l^6}
         \\&c_{1}^{(8)}=\frac{2(1+10n^2-15n^4+l^2(1+(-3+m)n^2))}{l^6}
     \end{split}
 \end{equation}

Finally, submitting the values of ($c_{1}^{(i)},c_{2}^{(i)}$) into (\ref{antz}), our solution can be finally written as (\ref{sol}).
 \section{Solution}
 \begin{equation}
     \begin{split}
         f_{0}=&\frac{12(4\alpha_{5}+\alpha_{6})}{r^3(r^2+n^2)^7}[-n^{12}+n^{10}(m^2-3mr-24r^2)+n^8r(3m^3+12m^2r-226mr^2\\&+192r^3)+2n^6r^3(19m^3-282m^2r+450mr^2-145r^3)+n^4r^5(-360m^3\\&+1054m^2r-702mr^2+129r^3)+n^2r^7(234m^3-285m^2r+95mr^2-6r^3)\\&-11m^3r^9+6m^2r^{10}]+\frac{8\alpha_{7}}{r^3(r^2+n^2)^7}[-27n^{12}+3n^{10}(9m^2-27mr-223r^2)\\&+3n^8r(27m^3+111m^2r-2116mr^2+1826r^3)+2n^6r^3(518m^3-7947m^2r \\&+12897mr^2-4195r^3)+3n^4r^5(-3430m^3+10226m^2r-6900mr^2+1287r^3)\\&+3n^2r^7(2324m^3-2871m^2r+973mr^2-63r^3)-343m^3r^9+189m^2r^{10}]
         \\&+\frac{2\alpha_{8}}{r^3(r^2+n^2)^7}[9n^{12}+n^{10}(-9m^2+27mr+174r^2)-3n^8r(9m^3+30m^2r\\&-514mr^2+380r^3)+3n^4r^5(700m^3-1682m^2r+942mr^2-135r^3)
         \\&-3mn^2r^7(266m^2-243mr+49r^2)+7m^3r^9]
     \end{split}
 \end{equation}
\begin{equation}
    \begin{split}
        h_{0}=&\frac{12(4\alpha_{5}+\alpha_{6})}{r(r^2+n^2)^7}[-n^{10}(m-3r)+n^8(m^3-15m^2r+54mr^2-46r^3)\\&+2n^6r^2(-9m^3+61m^2r-96mr^2+30r^3)+2n^4r^4(30m^3-96m^2r\\&+61mr^2-9r^3)+n^2r^6(-46m^3+54m^2r-15mr^2+r^3)+3m^3r^8-m^2r^9]\\&+\frac{8\alpha_{7}}{r(r^2+n^2)^7}[n^{10}(-27m+60r)+3n^8(9m^3-132m^2r+404mr^2-316r^3)\\&+2n^6r^2(-238m^3+1314m^2r-1845mr^2+530r^3)+6n^4r^4(175m^3-494m^2r
        \\&+258mr^2-18r^3)-3mn^2r^6(196m^2-180mr+17r^2)+35m^3r^8]
        \\&+\frac{2\alpha_{8}}{r(r^2+n^2)^7}[n^{10}(9m-69r)-3n^8(3m^3-51m^2r+326mr^2-334r^3)+2n^6r^2\\&\times(91m^3-1215m^2r+2358mr^2-830r^3)+6n^4r^4(-280m^3+1028m^2r-765mr^2\\&+153r^3)+3n^2r^6(574m^3 -774m^2r+281mr^2-21r^3)-119m^3r^8+63m^2r^9]
    \end{split}
\end{equation}
\begin{equation}
     \begin{split}
         f_{1}=&\frac{48\alpha_{5}+12\alpha_{6}}{r^2(n^2+r^2)^7}[-11n^{14}+3n^{12}(m^2-8mr-86r^2)+9n^{10}r^2(2m^2- 176mr\\&+239r^2)-3n^8r^4(701m^2-2400mr+1244r^2)+n^6r^6(4468m^2-6192 mr\\&+1611r^2)+3n^4r^8(-401m^2+168mr+18r^2)+n^2r^{10}(-102m^2+96mr-7r^2)\\&+7m^2r^{12}]+\frac{24\alpha_{7}}{7r^2(n^2+r^2)^7}[-99n^{14}+3n^{12}(9m^2-72mr-800r^2)+3 n^{10}r^2(56m^2\\&-4944mr+6811r^2)+n^8r^4(-19713m^2+68592mr-35896r^2)+n^6r^6(43240m^2\\&-60864mr+16191r^2)+n^4r^8(-12231m^2+5416mr+504r^2)-7n^2r^{10}(144m^2\\&-144mr+11r^2)+77 m^2 r^{12}]+\frac{6\alpha_{8}}{7r^2(n^2+r^2)^7}[33n^{14}-3n^{12}(3m^2-24mr-206r^2)\\&-3n^{10}r^2(14m^2-1200mr+1435r^2)+n^8r^4(4737m^2-14016mr+6580r^2)-n^6r^6\\&\times(7348m^2-8304mr+1449r^2)+n^4r^8(801m^2+248mr-126r^2)-7n^2r^{10}(-18m^2
         \\&+r^2)+7m^2r^{12}]
     \end{split}
 \end{equation}
\begin{equation}
    \begin{split}
        h_1=& \frac{192 n^2 \alpha_{5}+48 n^2 \alpha_6}{r(n^2+r^2)^7}[-n^{10} (2 m-9 r)-3 n^8 r (5 m^2-36 m r+46 r^2) +2 n^6 r^3 (61 m^2\\&-192 m r+90 r^2)
        -2 n^4 r^5 (96 m^2-122 m r+27 r^2)+3 n^2 r^7 (18 m^2-10 m r+r^2)\\&-m^2 r^9]+\frac{48 \alpha_7}{7r(n^2+r^2)^7}[-3 n^{12} (12 m
        -41 r)-3 n^{10} r (89 m^2-552 m r+649 r^2)
        \\&+n^8 r^3 (1803 m^2-5016 m r+2086 r^2)-2 n^6 r^5 (971 m^2
        -912 m r+63 r^2)+n^4 r^7\\&\times(270 m^2-100 m r+63 r^2)-n^2 r^9 (63 m^2-72 m r+7 r^2)+7 m^2 r^{11}]\\&+\frac{24 \alpha_8}{7r(n^2+r^2)^7}[6 n^{12} (m-11 r)+3 n^{10} r (16 m^2-204 m r+317 r^2)+n^8 r^3 (-759 m^2\\&+3048 m r-1694 r^2)
        +2 n^6 r^5 (1045 m^2-1650 m r+504 r^2)+2 m n^4 r^7 (265 r-432 m)
        \\&+n^2 r^9 (-42 m^2+72 m r-7 r^2)+7m^2 r^{11}]
    \end{split}
\end{equation}
\begin{equation}
    \begin{split}
        f_2=&\frac{384 n^4 \alpha_{5}+96n^4 \alpha_{6}}{r^2(n^2+r^2)^7}[-5n^{12}-6 n^{10} r (m+19 r)+n^8 r^3 (999 r-340 m)+24 n^6 r^5\\&\times(75 m-83 r)-9 n^4 r^7 (188 m
        -93 r)+2 n^2 r^9 (31 m+45 r) +(48 m-7 r)r^{11}]\\&+\frac{192 n^4 \alpha_7}{7r^2(n^2+r^2)^7}
        [-45n^{12}-18 n^{10} r (3 m+59 r)+n^8 r^3 (9477 r-3184 m)+36 n^6 r^5 \\&\times(475 m-531 r)
        -9 n^4 r^7 (1848 m-941 r)+6 n^2 r^9 (127 m+147 r)+7 r^{11} (72 m\\&-11 r)]+\frac{48 n^4 \alpha_8}{7r^2(n^2+r^2)^7}
        [15 n^{12}+18 n^{10} r (m+15 r)+n^8 r^3 (772 m-2025 r)\\&-3600 n^6 r^5 (m-r)+9 n^4 r^7 (252 m-71 r)+6 n^2 r^9 (37 m-21 r)-7 r^{12}]
    \end{split}
\end{equation}
\begin{equation}
    \begin{split}
        h_2=&\frac{-768 n^6\alpha_{5}-192n^6 \alpha_{6}}{r(n^2+r^2)^7}[m  (n^8-54 n^6 r^2+192 n^4 r^4-122 n^2 r^6+15 r^8 )\\&-3(3 n^8 r-46n^6 r^3+60 n^4 r^5-18 n^2 r^7+r^9 )]+\frac{-384 n^4 \alpha_7}{7r(n^2+r^2)^7}[m (9 n^{10}
        \\&-424 n^8 r^2+1278 n^6 r^4-396 n^4 r^6+33 n^2 r^8-36 r^{10})-63 n^{10} r+999 n^8 r^3\\&-1026 n^6 r^5+18 n^4 r^7-63 n^2 r^9+7 r^{11}]+\frac{96 n^4 \alpha_8}{7r(n^2+r^2)^7}[3 n^{10} (m-21 r)\\&+n^8 (900 r^3-286 m r^2)+36 n^6 r^4 (41 m-48 r)-6n^4 r^6 (295 m-183 r)+3 n^2 r^8\\&\times(83 m+21 r)+72 m r^{10}-14 r^{11}]
    \end{split}
\end{equation}
\begin{equation}
    \begin{split}
        f_3=&\frac{144\alpha_1+36\alpha_2+9\alpha_3+9\alpha_4}{n^2+r^2}(-3 n^4+6 n^2 r^2+r^4)+\frac{-24  \alpha_{5}-6 \alpha_{6}}{r^2(n^2+r^2)^7}(96 n^{18}\\&+2115 n^{16} r^2
        -19668 n^{14} r^4+45192 n^{12} r^6-18564 n^{10} r^8-3354 n^8 r^{10}+132 n^6 r^{12}\\&-48 n^4 r^{14}-12 n^2 r^{16}-r^{18})+\frac{-4\alpha_7}{7r^2(n^2+r^2)^7}(2592 n^{18}+59157 n^{16} r^2-559116 n^{14} r^4\\&+1298232 n^{12} r^6-566172 n^{10} r^8-97398 n^8 r^{10}+6748 n^6 r^{12}-336 n^4 r^{14}-84 n^2 r^{16}\\&-7r^{18})+\frac{2\alpha_8}{7r^2(n^2+r^2)^7}
        (432 n^{18}+7371 n^{16} r^2-60804 n^{14} r^4+125064 n^{12} r^6\\&-17460 n^{10} r^8-5418 n^8 r^{10}+308 n^6 r^{12}+336 n^4 r^{14}+84 n^2 r^{16}+7 r^{18})
    \end{split}
\end{equation}
\begin{equation}
    \begin{split}
        h_3=&\frac{144\alpha_1+36\alpha_2+9\alpha_3+9\alpha_4}{n^2+r^2}(-3 n^4+6 n^2 r^2+r^4)+\frac{24  \alpha_{5}+6 \alpha_{6}}{(n^2+r^2)^7}(381 n^{16}-5900 n^{14} r^2\\&
        +7672 n^{12} r^4-2268 n^{10} r^6+218 n^8 r^8+92 n^6 r^{10}+48 n^4 r^{12}+12 n^2 r^{14}+r^{16})\\&
        +\frac{4\alpha_7}{7(n^2+r^2)^7}(8235 n^{16}-131220 n^{14} r^2+129096 n^{12} r^4+3708 n^{10} r^6+12726 n^8 r^8\\&
        -700 n^6 r^{10}+336 n^4 r^{12}+84 n^2 r^{14}+7 r^{16})+\frac{2\alpha_8}{7(n^2+r^2)^7}(-3861 n^{16}+54252 n^{14} r^2\\ &
        -112824 n^{12} r^4+76284 n^{10} r^6+8694 n^8 r^8-700 n^6 r^{10}+336 n^4 r^{12}+84 n^2 r^{14}+7 r^{16})
    \end{split}
\end{equation}

\end{document}